\documentclass[fleqn,usenatbib]{mnras}

\usepackage{newtxtext,newtxmath}

\usepackage[T1]{fontenc}

\DeclareRobustCommand{\VAN}[3]{#2}
\let\VANthebibliography\thebibliography
\def\thebibliography{\DeclareRobustCommand{\VAN}[3]{##3}\VANthebibliography}

\usepackage{graphicx}	
\usepackage{amsmath}	
\usepackage{tabularx}

\newcommand{\sfive}{S5\,0716+714}	
\newcommand{\bl}{BL\,Lacertae}

\title[Multi-wavelength variability of S5 0716+714]{Multi-wavelength variability analysis of the blazar S5 0716+714 during a long-lasting period of low activity}

\author[B. M. Mihov et al.]{
B. M. Mihov,$^{1}$\thanks{E-mail: bmihov@astro.bas.bg (BMM)}
E. G. Elhosseiny,$^{2}$
L. S. Slavcheva-Mihova,$^{1}$
Ali Takey,$^{2}$
M. N. Ismail$^{3}$
and Ramy Mawad$^{3}$
\\
$^{1}$Institute of Astronomy and NAO, Bulgarian Academy of Sciences, 72 Tsarigradsko Chaussee Blvd., 1784 Sofia, Bulgaria \\
$^{2}$National Research Institute of Astronomy and Geophysics (NRIAG), 11421 Helwan, Cairo, Egypt \\
$^{3}$Astronomy \& Meteorology Department, Faculty of Science, Al-Azhar University, 11488 Cairo, Egypt
}

\date{Accepted XXX. Received YYY; in original form ZZZ}

\pubyear{\the\year{}}

\begin{document}
\label{firstpage}
\pagerange{\pageref{firstpage}--\pageref{lastpage}}
\maketitle

\begin{abstract}

We conducted a multi-wavelength monitoring campaign of the blazar S5 0716+714 from 2022 November 26 to 2023 May 28 using optical telescopes in Egypt and Bulgaria. Data were taken during 84 nights in 11 of which intranight monitoring was performed. We also use optical and $\gamma$-ray survey data. On long-term time-scales, we find a gradual decrease of the S5 0716+714 activity since the beginning of 2020 in both optical and $\gamma$-rays. 
On short-term time-scales, the individual optical light curves are strongly correlated among each other with no time lags observed. 
The $V$-band percentage variability amplitude equals $97.59 \pm 0.02$ per cent.
We find moderate flatter-when-brighter spectral behaviour with the strength of the `spectral index~-- flux' anti-correlation decreasing towards the longer wavelengths. The main feature of the short-term light curves is a transient quasi-periodic oscillation with a period of $43.5 \pm 3.6$\,d. The $V$-band light curve is modelled with two helically moving blobs and a synchrotron flare. We estimate the resulting parameters, as well as limits on the radius, magnetic field strength, and electron Lorentz factor of the region responsible for the flare. On intranight time-scales, we find smooth flux variations with no flares and derive a duty cycle in the range $\sim$10--20 per cent. The lack of flares on intranight time-scales could result from a temporarily homogeneous jet flow without formation of turbulent cells in terms of prevented Kelvin-Helmholtz instability. The analysis of the data reveals a low activity of S5 0716+714 on all time-scales during the observation period.

\end{abstract}




\section{Introduction}

Blazars are a subclass of radio-loud active galactic nuclei whose jets are closely aligned with the line-of-sight towards the observer \citep{1980ARA&A..18..321A}. One of the main property of blazars is their violent variability across the entire electromagnetic spectrum~-- in the optical the amplitudes are up to several magnitudes \citep[e.g.][]{1993A&A...278..391S,2006A&A...453..817V,2017Natur.552..374R,2018BlgAJ..28...22B,2022Natur.609..265J}. 

Blazars could broadly be divided into \bl\ objects (BL\,Lacs) and flat-spectrum radio quasars (FSRQs) depending on the shape of their optical spectra: FSRQs show broad emission lines typical for quasars, while BL\,Lacs show featureless continuum or emission lines with equivalent width less than or equal to 5\,\AA.

The blazar spectral energy distribution (SED) consists of two humps~-- one from radio to X-rays and another from X- to $\gamma$-rays. The first hump results from the synchrotron emission of relativistic electrons, while the second one \citep[within the framework of leptonic scenario;][]{2013ApJ...768...54B} is due to the inverse Compton scattering of low-energy seed photons by the same relativistic electrons. The position of the synchrotron peak is used to classify blazars as follows \citep{2010ApJ...716...30A}: low synchrotron peaked blazars (the synchrotron hump peaks at frequencies less than $10^{14}$\,Hz), intermediate synchrotron peaked blazars (the synchrotron hump peaks at frequencies $10^{14} - 10^{15}$\,Hz), and high synchrotron peaked blazars (the synchrotron hump peaks at frequencies greater than $10^{15}$\,Hz).

Considering the time-scales, the blazar variability could be divided into: long-term variability (LTV), short-term variability (STV), or intranight variability (INV or intraday variability or microvariability) when the time-scale ranges from months to years/decades, from days to weeks/months, or from minutes to hours \citep[but less than a day;][]{2020AN....341..713S}, respectively. Observed time-scales could be transformed back to the rest-frame via $t' = t\delta/(1+z)$, where $\delta$ is the Doppler factor and $z$ is the redshift; throughout the paper the primed quantities are in the rest-frame.

The BL\,Lac \sfive\ is among the best studied blazars. It is classified as an intermediate synchrotron peaked blazar \citep{2010ApJ...716...30A,2016ApJS..226...20F,2022ApJS..262...18Y}. Its redshift is still uncertain owing to its featureless optical spectrum \citep{1993A&AS...98..393S,1996MNRAS.281..425M}. Recently, \citet{2023A&A...680A..52P} concluded with a probability greater than 99 per cent that \sfive\ belongs to a group of galaxies. The mean redshift of the group, and, hence, of \sfive, is $0.2304 \pm 0.0013$.

\sfive\ is believed to have a high duty cycle (DC) on intranight time-scales throughout the years \citep[e.g.][]{1995ARA&A..33..163W,2009ApJS..185..511P,2016MNRAS.455..680A,2016Galax...4...15W,2019Ap&SS.364...83W}. A recent confirmation of the high DC of \sfive\ comes from the study of \citet{2024MNRAS.527.5220T}. The authors derived $BVRI$-band DC values of 58--75 per cent that rose up to 80--85 per cent when accounting also for literature intranight light curves (LCs) over the last three decades. Furthermore, \citet{2024ApJ...971...74E} reported a DC of 64 per cent, while the data of \citet{2025ApJS..277...18L} suggest a DC value in the range 39--72 per cent.
Low DC values ($\lesssim$30 per cent) were reported as well \citep{2015ApJS..218...18D,2018AJ....155...31H,2018AJ....156...36K}.

The colour (or spectral) behaviour of \sfive\ is generally reported to be bluer-when-brighter (BWB, or flatter-when-brighter, FWB, if we consider the spectral index instead of the colour index). The redder-when-brighter (or steeper-when-brighter) chromatism is typical for FSRQs. In particular, \citet{2021MNRAS.501.1100R} reported dependence of the \sfive\ colour behaviour on the time-scale considered~-- the short-term flux variations are mildly chromatic, while the rapid, intranight flux changes are strongly chromatic \citep[see also][]{2014MNRAS.443.2940H}. The same behaviour was noticed for \bl\ \citep[e.g.][]{2002A&A...390..407V,2004A&A...421..103V,2023ApJS..265...51A}. Another interesting feature in the \sfive\ spectral behaviour is the spectral index saturation as the flux increases. At low brightness states a strong FWB chromatism is observed, while at high states the FWB trend becomes insignificant \citep{2020ApJS..247...49X,2020ApJ...902...42F,2022ApJ...928...86G}. The spectral index flattening at high states is observed in other blazars as well \citep[e.g.][]{2022ApJS..259...49Z}. The authors proposed a more general colour behaviour classification, namely bluer-stable-when-brighter and redder-stable-when-brighter. \citet{2017A&A...605A..43Y} and \citet{2021RAA....21..102L} reported that the strength of the anti-correlation between the spectral index and flux depends on the band: the anti-correlation gets stronger towards the shorter wavelengths.

Quasi-periodic oscillations (QPOs) of \sfive\ in the optical were reported on long-term to intranight time-scales with periods $1060 \pm 50$\,d \citep{2023Ap&SS.368...88H}, $44 \pm 6$\,d \citep{2024ApJ...961..180L}, $\sim$50\,min \citep{2018AJ....155...31H}, and $\sim$15\,min \citep{2010ApJ...719L.153R}. The QPOs were also detected in the radio and $\gamma$-rays \citep[e.g.][]{2018Ap&SS.363..169L,2022ApJ...938....8C,2023ApJ...943..157L}.

In this paper, we present the results from a multi-wavelength (MWL) monitoring campaign of \sfive\ on short-term and intranight time-scales using optical telescopes in Egypt and Bulgaria. We also analyse optical and $\gamma$-ray survey data on long-term time-scales.

The paper is organized as follows. In Sect.\,\ref{sec:obs} we present the observations and data reduction; the survey data are depicted in Sect.\,\ref{sec:asas}. In Sect.\,\ref{sec:photo} we describe the photometric techniques used, while in Sects.\,\ref{sec:res} and \ref{sec:disc} we present and discuss the results, respectively. Finally, a short summary of the results is given in Sect.\,\ref{sec:sum}. 

\begin{table*}
    \centering
    \caption{Parameters of the CCD cameras used in our observations.}
    \label{tab:tels}
    \begin{tabular}{ccccccc}
    \hline
    Telescope & CCD model & Chip size & CCD scale & Field-of-view & Gain & Read-out noise \\
    &  & [px] & [arcsec px$^{-1}$] & [arcmin] & [e$^{-}$ ADU$^{-1}$] & [e$^{-}$ px$^{-1}$ RMS] \\
    \hline
    Kot 188 cm & E2V CCD42-40 & 2048$\times$2048 & 0.238 & 8.10$\times$8.10 & 2.14 & 3.92 \\
    Roz 200 cm & Andor iKon-L & 2048$\times$2048 & 0.497 & 17.0$\times$17.0 & 1.0$^{(a)}$/1.1$^{(b)}$ & 6.7$^{(a)}$/6.9$^{(b)}$ \\
    \hline
    \end{tabular} \\
    \flushleft{{\em Note.} $^{(a)}$Parameters of the CCD camera at the FoReRo-2 blue channel. $^{(b)}$Parameters of the CCD camera at the FoReRo-2 red channel.}
\end{table*}

\begin{table*}
    \centering
    \caption{Log of the \sfive\ observations.}
    \label{tab:log}
    \resizebox{\textwidth}{!}{
    \begin{tabular}{@{}cccccccccccccccccccc@{}}
    \hline
    Date & Telescope & \multicolumn{8}{c}{Number of data points} & Date & Telescope & \multicolumn{8}{c}{Number of data points} \\
    \cline{3-10}\cline{13-20}\noalign{\smallskip}
     & & $B$ & $V$ & $R$ & $I$ & $g$ & $r$ & $i$ & $z$ & & & $B$ & $V$ & $R$ & $I$ & $g$ & $r$ & $i$ & $z$ \\
    \hline
    2022 Nov 26  & Kot 188 cm & 0  &  0  &  5  &  0  &  0  &  0  & 0  &  0  & 2023 Feb 16  & Kot 188 cm & 3  &  3  &  3  &  0  &  0  &  0  & 0  &  0  \\
    2022 Nov 28  & Kot 188 cm & 3  &  3  &  3  &  0  &  0  &  0  & 0  &  0  & 2023 Feb 17  & Kot 188 cm & 3  &  3  &  3  &  0  &  0  &  0  & 0  &  0  \\  
    2022 Nov 29  & Kot 188 cm & 6  &  4  &  3  &  0  &  0  &  0  & 0  &  0  & 2023 Feb 18  & Kot 188 cm & 3  &  3  & 11  &  0  &  1  &  0  &490 &  0  \\  
    2022 Nov 30  & Kot 188 cm & 5  &  3  &  3  &  0  &  0  &  0  & 0  &  0  & 2023 Feb 19  & Kot 188 cm & 2  &  0  &  0  &  0  &  0  &  0  & 0  &  0  \\  
    2022 Dec 01  & Kot 188 cm & 3  &  3  &  3  &  0  &  0  &  0  & 0  &  0  & 2023 Feb 21  & Kot 188 cm & 1  &  7  &  8  &  0  &  0  &  0  & 0  &  0  \\  
    2022 Dec 05  & Kot 188 cm & 10 & 12  &  12 &  0  &  0  &  0  & 0  &  0  & 2023 Feb 22  & Kot 188 cm & 4  &  5  &  5  &  0  &  0  &  0  & 0  &  0  \\  
    2022 Dec 06  & Kot 188 cm & 3  &  3  &  3  &  0  &  0  &  0  & 0  &  0  & 2023 Feb 24  & Kot 188 cm & 2  &  3  &  3  &  0  &  0  &  0  & 0  &  0  \\  
    2022 Dec 07  & Kot 188 cm & 3  &  3  &  3  &  0  &  0  &  0  & 0  &  0  & 2023 Feb 25  & Kot 188 cm & 1  &  3  &  3  &  0  &  0  &  0  & 0  &  0  \\  
    2022 Dec 12  & Kot 188 cm & 5  &  5  &  5  &  0  &  0  &  0  & 0  &  0  & 2023 Feb 26  & Kot 188 cm & 2  &  3  &  3  &  0  &  0  &  0  & 0  &  0  \\  
    2022 Dec 13  & Kot 188 cm & 5  &  5  &  5  &  0  &  0  &  0  & 0  &  0  & 2023 Mar 09  & Kot 188 cm & 6  &  6  &  6  &  0  &  0  &  0  & 0  &  0  \\  
    2022 Dec 14  & Kot 188 cm & 10 & 10  &  10 &  0  &  0  &  0  & 0  &  0  & 2023 Mar 10  & Kot 188 cm & 3  &  3  &  3  &  0  &  0  &  0  & 0  &  0  \\  
    2022 Dec 15  & Kot 188 cm & 10 & 10  &  10 &  0  &  0  &  0  & 0  &  0  & 2023 Mar 11  & Kot 188 cm & 3  &  3  &  4  &  0  &  0  &  0  & 0  &  0  \\  
    2022 Dec 17  & Kot 188 cm & 5  &  5  &  5  &  0  &  0  &  0  & 0  &  0  & 2023 Mar 12  & Kot 188 cm & 3  &  3  &  3  &  0  &  0  &  0  & 0  &  0  \\  
    2022 Dec 18  & Kot 188 cm & 5  &  5  &  5  &  0  &  0  &  0  & 0  &  0  & 2023 Mar 14  & Kot 188 cm & 3  &  3  &  3  &  0  &  0  &  0  & 0  &  0  \\  
    2022 Dec 21  & Roz 200 cm & 378&  0  &  0  & 1014&  0  &  0  & 0  &  0  & 2023 Mar 17  & Kot 188 cm & 3  &  3  &  3  &  0  &  47 &  0  & 46 &  0  \\  
    2022 Dec 24  & Kot 188 cm & 3  &  3  &  10 &  0  &  0  &  0  & 0  &  0  & 2023 Mar 18  & Kot 188 cm & 3  &  3  &  3  &  0  &  0  &  0  & 0  &  0  \\  
    2022 Dec 26  & Kot 188 cm & 3  &  3  &  3  &  0  & 501 &  0  &502 &  0  & 2023 Mar 19  & Kot 188 cm & 3  &  3  &  8  &  0  & 240 &  0  &240 &  0  \\  
    2022 Dec 27  & Kot 188 cm & 3  &  3  &  3  &  0  & 497 &  0  &495 &  0  & 2023 Mar 21  & Kot 188 cm & 5  &  5  &  5  &  0  &  0  &  0  & 0  &  0  \\  
    2022 Dec 28  & Kot 188 cm & 3  &  3  &  3  &  0  & 526 &  0  &542 &  0  & 2023 Mar 25  & Kot 188 cm & 3  &  2  &  3  &  0  &  0  &  0  & 0  &  0  \\  
    2022 Dec 29  & Kot 188 cm & 3  &  3  &  3  &  0  & 495 &  0  &495 &  0  & 2023 Mar 27  & Kot 188 cm & 4  &  5  &  5  &  0  &  0  &  0  & 0  &  0  \\  
    2023 Jan 02  & Kot 188 cm & 25 & 24  & 23  &  0  &  0  &  0  & 0  &  0  & 2023 Mar 28  & Kot 188 cm & 2  &  3  &  3  &  0  &  0  &  0  & 0  &  0  \\  
    2023 Jan 04  & Kot 188 cm & 8  & 10  & 10  &  0  &  0  &  0  & 0  &  0  & 2023 Mar 30  & Kot 188 cm & 2  &  3  &  3  &  0  &  0  &  0  & 0  &  0  \\  
    2023 Jan 09  & Kot 188 cm & 21 & 21  & 21  &  0  &  0  &  0  & 0  &  0  & 2023 Apr 02  & Kot 188 cm & 3  &  3  &  3  &  0  &  0  &  0  & 0  &  0  \\  
    2023 Jan 10  & Kot 188 cm & 11 & 15  & 16  &  0  &  0  &  0  & 0  &  0  & 2023 Apr 06  & Kot 188 cm & 3  &  3  &  6  &  0  &  0  &  0  & 0  &  0  \\  
    2023 Jan 11  & Kot 188 cm & 12 & 19  & 15  &  0  &  0  &  0  & 0  &  0  & 2023 Apr 07  & Kot 188 cm & 3  &  3  &  8  &  0  &  0  &  0  & 0  &  0  \\  
    2023 Jan 12  & Kot 188 cm & 5  &  5  &  8  &  0  &  0  &  0  & 0  &  0  & 2023 Apr 10  & Kot 188 cm & 3  &  3  &  3  &  0  &  0  &  0  & 0  &  0  \\  
    2023 Jan 13  & Kot 188 cm & 2  &  5  &  5  &  0  &  0  &  0  & 0  &  0  & 2023 Apr 12  & Kot 188 cm & 2  &  2  &  3  &  0  &  0  &  0  & 0  &  0  \\  
    2023 Jan 16  & Kot 188 cm & 5  &  5  &  5  &  0  &  0  &  0  & 0  &  0  & 2023 Apr 13  & Kot 188 cm & 3  &  3  &  3  &  0  &  0  &  0  & 0  &  0  \\  
    2023 Jan 17  & Kot 188 cm & 5  &  5  &  5  &  0  &  0  &  0  & 0  &  0  & 2023 Apr 14  & Kot 188 cm & 3  &  3  &  3  &  0  &  0  &  0  & 0  &  0  \\  
    2023 Jan 18  & Kot 188 cm & 4  &  5  &  5  &  0  &  0  &  0  & 0  &  0  & 2023 Apr 25  & Kot 188 cm & 3  &  5  &  5  &  0  &  0  &  0  & 0  &  0  \\  
    2023 Jan 19  & Kot 188 cm & 5  &  5  &  5  &  0  &  0  &  0  & 0  &  0  & 2023 Apr 28  & Kot 188 cm & 3  &  3  &  3  &  0  &  0  &  0  & 0  &  0  \\  
    2023 Jan 20  & Kot 188 cm & 5  &  5  &  5  &  0  &  0  &  0  & 0  &  0  & 2023 Apr 29  & Kot 188 cm & 3  &  3  &  3  &  0  &  0  &  0  & 0  &  0  \\  
    2023 Jan 21  & Kot 188 cm & 4  &  5  &  5  &  0  &  0  &  0  & 0  &  0  & 2023 Apr 30  & Kot 188 cm & 3  &  3  &  3  &  0  &  0  &  0  & 0  &  0  \\  
    2023 Jan 22  & Kot 188 cm & 5  &  5  &  5  &  0  &  0  &  0  & 0  &  0  & 2023 May 08  & Kot 188 cm & 3  &  3  &  3  &  0  &  0  &  0  & 0  &  0  \\  
    2023 Jan 25  & Kot 188 cm & 3  &  3  &  3  &  0  & 258 & 257 & 257&257  & 2023 May 09  & Kot 188 cm & 3  &  3  &  3  &  0  &  0  &  0  & 0  &  0  \\  
    2023 Jan 26  & Kot 188 cm & 3  &  3  &  3  &  0  &  0  &  0  & 0  &  0  & 2023 May 12  & Kot 188 cm & 3  &  3  &  3  &  0  &  0  &  0  & 0  &  0  \\  
    2023 Jan 28  & Kot 188 cm & 2  &  6  &  6  &  0  &  0  &  0  & 0  &  0  & 2023 May 13  & Kot 188 cm & 3  &  3  &  3  &  0  &  0  &  0  & 0  &  0  \\  
    2023 Jan 29  & Kot 188 cm & 2  &  3  &  3  &  0  &  0  &  0  & 0  &  0  & 2023 May 17  & Kot 188 cm & 3  &  3  &  3  &  0  &  0  &  0  & 0  &  0  \\  
    2023 Feb 07  & Kot 188 cm & 1  &  3  &  3  &  0  &  0  &  0  & 0  &  0  & 2023 May 22  & Kot 188 cm & 2  &  1  &  3  &  0  &  0  &  0  & 0  &  0  \\  
    2023 Feb 09  & Kot 188 cm & 3  &  3  &  3  &  0  & 502 &  0  & 505&  0  & 2023 May 23  & Kot 188 cm & 3  &  2  &  3  &  0  &  0  &  0  & 0  &  0  \\  
    2023 Feb 10  & Kot 188 cm & 3  &  3  &  3  &  0  &  0  &  0  & 0  &  0  & 2023 May 25  & Kot 188 cm & 3  &  3  &  3  &  0  &  0  &  0  & 0  &  0  \\ 
    2023 Feb 14  & Roz 200 cm & 0  &  0  &  0  &  0  & 612 &  0  & 0  & 682 & 2023 May 28  & Kot 188 cm & 6  &  6  &  6  &  0  &  0  &  0  & 0  &  0  \\ 
    \hline
    \end{tabular}}
\end{table*}

\begin{table*}
    \centering
    \caption{Intranight monitoring characteristics and results.}
    \label{tab:inv}
    \resizebox{\textwidth}{!}{
    \begin{tabular}{@{}ccccccccccc@{}}
    \hline
    Date & Band & Integr. & Sampl. & Durat. & $\langle \rm FWHM \rangle_{med}$ & $\langle m_{\rm blz} \rangle_{\rm wt}$ & $\langle m_{\rm ctrl} \rangle_{\rm wt}$ & $\kappa$ & {$C_{\kappa}$, Status} & ${\cal A}$ \\
    & & [s] & [s] & [h] & [arcsec] & [mag] & [mag] & & & [per cent] \\
    (1) & (2) & (3) & (4) & (5) & (6) & (7) & (8) & (9) & (10) & (11) \\
    \hline
    2022 Dec 21 & $B$ & 30 & 33.7 & 3.91 & 1.35, 0.10 & 14.564, 2.4& 14.225, 0.3& 1.160 & 2.155, NV     & 4.175 $\pm$ 0.010 \\
                &     &    &      &      &            & 0.001, 0.010& 0.001, 0.004&       &               &                   \\
                & $I$ & 10 & 13.8 & 3.94 & 1.98, 0.27 & 13.196, 2.6& 12.965, 2.2& 1.179 & 1.018~~~~~~~~~& 3.921 $\pm$ 0.005 \\
                &     &    &      &      &            & 0.001, 0.006& 0.001, 0.005&       &               &                   \\
    \noalign{\smallskip}
    2022 Dec 26 & $g$ & 5  & 30.2 & 4.30 & 2.43, 0.36 & 14.261, 0.4& 13.864, 0.2& 1.341 & 1.119, NV     & 5.286 $\pm$ 0.022 \\
                &     &    &      &      &            & 0.001, 0.009& 0.001, 0.006&       &               &                   \\
	            & $i$ & 2  & 30.2 & 4.27 & 2.31, 0.42 & 13.754, 0.2& 13.412, 0.1& 1.196 & 1.075~~~~~~~~~& 5.596 $\pm$ 0.036 \\
                &     &    &      &      &            & 0.001, 0.009& 0.001, 0.007&       &               &                   \\
    \noalign{\smallskip}
    2022 Dec 27 & $g$ & 7  & 33.0 & 4.67 & 1.91, 0.41 & 14.146, 0.6& 13.864, 0.1& 1.223 & 1.962, NV     & 4.594 $\pm$ 0.022 \\
                &     &    &      &      &            & 0.001, 0.012& 0.001, 0.005&       &               &                   \\
	            & $i$ & 4  & 33.0 & 4.70 & 1.84, 0.36 & 13.655, 0.3& 13.410, 0.1& 1.140 & 1.930~~~~~~~~~& 4.452 $\pm$ 0.036 \\	    
                &     &    &      &      &            & 0.001, 0.011& 0.001, 0.005&       &               &                   \\
    \noalign{\smallskip}
    2022 Dec 28 & $g$ & 7  & 32.7 & 4.91 & 2.03, 0.32 & 14.184, 0.1& 13.863, 0.1& 1.255 & 0.996, NV     & 2.057 $\pm$ 0.028 \\
                &     &    &      &      &            & 0.001, 0.005& 0.001, 0.004&       &               &                   \\
    	        & $i$ & 4  & 32.7 & 4.91 & 1.94, 0.33 & 13.675, 0.1& 13.408, 0.1& 1.150 & 0.870~~~~~~~~~&        0.0        \\
                &     &    &      &      &            & 0.001, 0.005& 0.001, 0.005&       &               &                   \\
    \noalign{\smallskip}
    2022 Dec 29 & $g$ & 7  & 32.2 & 4.50 & 1.79, 0.29 & 14.184, 0.1& 13.865, 0.1& 1.254 & 1.196, NV     & 2.492 $\pm$ 0.026 \\
                &     &    &      &      &            & 0.001, 0.006& 0.001, 0.004&       &               &                   \\
    	        & $i$ & 4  & 32.6 & 4.50 & 1.71, 0.29 & 13.689, 0.1& 13.410, 0.1& 1.157 & 0.864~~~~~~~~~& 0.771 $\pm$ 0.122 \\	
                &     &    &      &      &            & 0.001, 0.005& 0.001, 0.005&       &               &                   \\
    \noalign{\smallskip}
    2023 Jan 25 & $g$ & 12 & 75.4 & 5.58 & 2.13, 0.52 & 14.593, 0.4& 13.864, 0.1& 1.836 & 0.980, NV     & 4.316 $\pm$ 0.022 \\
                &     &    &      &      &            & 0.001, 0.009& 0.001, 0.005&       &               &                   \\
    	          & $r$ & 10 & 75.4 & 5.58 & 2.09, 0.51 & 14.222, 0.1& 13.474, 0.1& 1.838 & 0.871~~~~~~~~~& 2.321 $\pm$ 0.049 \\
                &     &    &      &      &            & 0.001, 0.008& 0.001, 0.005&       &               &                   \\
    	          & $i$ & 8  & 75.4 & 5.58 & 2.15, 0.55 & 14.008, 0.1& 13.414, 0.1& 1.445 & 1.038~~~~~~~~~& 0.869 $\pm$ 0.108 \\	  
                &     &    &      &      &            & 0.001, 0.006& 0.001, 0.004&       &               &                   \\
    	        & $z$ & 10 & 75.4 & 5.58 & 1.99, 0.51 & 13.828, 0.1& 13.385, 0.1& 1.378 & 1.016~~~~~~~~~& 1.551 $\pm$ 0.066 \\
                &     &    &      &      &            & 0.001, 0.007& 0.001, 0.005&       &               &                   \\
    \noalign{\smallskip}
    2023 Feb 09 & $g$ & 12 & 42.4 & 5.85 & 2.44, 0.35 & 13.940, 0.2& 13.858, 0.1& 1.055 & 1.137, NV     & 2.676 $\pm$ 0.026 \\
                &     &    &      &      &            & 0.001, 0.006& 0.001, 0.005&       &               &                   \\
    	        & $i$ & 8  & 42.3 & 5.84 & 2.36, 0.34 & 13.407, 0.1& 13.407, 0.1& 0.985 & 1.184~~~~~~~~~& 3.697 $\pm$ 0.039 \\
                &     &    &      &      &            & 0.001, 0.007& 0.001, 0.006&       &               &                   \\
    \noalign{\smallskip}
    2023 Feb 14 & $g$ & 12 & 15.6 & 2.61 & 1.86, 0.48 & 13.734, 0.2& 13.858, 0.1& 0.955 & 2.094, NV     & 2.582 $\pm$ 0.026 \\
                &     &    &      &      &            & 0.001, 0.006& 0.001, 0.003&       &               &                   \\
                & $z$ &8,15& 12.1 & 2.61 & 1.47, 0.33 & 13.014, 0.1& 13.382, 0.1& 0.812 & 1.970~~~~~~~~~& 3.166 $\pm$ 0.042 \\
                &     &    &      &      &            & 0.001, 0.008& 0.001, 0.005&       &               &                   \\
    \noalign{\smallskip}
    2023 Feb 18 & $i$ & 20 & 23.0 & 4.80 & 2.74, 1.26 & 13.388, 1.3& 13.415, 0.1& 0.985 & 6.091, Var    & 6.806 $\pm$ 0.033 \\
                &     &    &      &      &            & 0.001, 0.024& 0.001, 0.004&       &               &                   \\
    \noalign{\smallskip}
    2023 Mar 17 & $g$ & 25 & 61.5 & 0.78 & 3.01, 1.13 & 14.376, 0.3& 13.864, 0.1& 1.569 & 1.147, NV     & 5.675 $\pm$ 0.026 \\
                &     &    &      &      &            & 0.002, 0.009& 0.002, 0.005&       &               &                   \\
	            & $i$ & 15 & 61.5 & 0.77 & 2.95, 1.11 & 13.794, 0.0& 13.412, 0.1& 1.355 & 0.923~~~~~~~~~&        0.0        \\	    
                &     &    &      &      &            & 0.003, 0.005& 0.003, 0.004&       &               &                   \\
    \noalign{\smallskip}
    2023 Mar 19 & $g$ & 30 & 71.6 & 4.76 & 2.75, 0.72 & 14.263, 0.8& 13.866, 0.1& 1.411 & 2.303, PV     & 6.265 $\pm$ 0.021 \\
                &     &    &      &      &            & 0.001, 0.013& 0.001, 0.004&       &               &                   \\
    	        & $i$ & 20 & 71.6 & 4.76 & 2.74, 0.85 & 13.697, 0.4& 13.414, 0.1& 1.231 & 3.791~~~~~~~~~& 4.470 $\pm$ 0.036 \\
                &     &    &      &      &            & 0.001, 0.014& 0.001, 0.003&       &               &                   \\
    \hline
    \end{tabular}}
    \flushleft{{\em Note.} Table columns read: Col.~3: integration time; Col.~4: median time sampling of the LCs; Col.~5: duration of the INM sessions; Col.~6: median FWHM during the INM sessions and its standard deviation; Col.~7: first row~-- the weighted mean magnitude of the blazar and reduced $\chi^2$. Second row~-- the standard uncertainty of the weighted mean and the weighted standard deviation; Col.~8: same as in Col.~7, but for the control star; Col.~9: scaling factor; Col.~10: scaled $C$-criterion and the variability status assigned for the night: Var~-- variable, PV~-- probably variable, NV~-- non-variable; Col.~11: percentage variability amplitude (defined in Sect.~\ref{sec:photo}). If the radicand in the ${\cal A}$ expression gets negative, then ${\cal A}$ is set to zero}.
\end{table*}

\section{Observations and data reduction} 
\label{sec:obs}

The blazar \sfive\ was observed from 2022 November 26 to 2023 May 28 by means of two ground-based optical telescopes: the 1.88\,m Cassegrain telescope at the Kottamia Astronomical Observatory (KAO, denoted as `Kot 188 cm'), Egypt, and  the 2\,m Ritchey-Chr\'etien telescope at the Rozhen National Astronomical Observatory (NAO, denoted as `Roz 200 cm'), Bulgaria. Both telescopes were used in combination with multi-mode focal reducers attached at their Cassegrain/Ritchey-Chr\'etien foci, namely the single-channel Kottamia Faint Imaging Spectro-Polarimeter \citep[KFISP;][]{2022ExA....53...45A} and
the two-channel Focal Reducer Rozhen \citep[FoReRo-2;][]{2000KFNTS...3...13J}. 
The CCD cameras used are described in Table~\ref{tab:tels}.

The data were obtained through the Johnson-Cousins $BVRI$ and Sloan $g'r'i'z'$ set of filters (we shall skip the primes throughout the paper for short). We generally took several frames per filter for the nights with no intranight monitoring (INM) performed. 
The log of the \sfive\ observations is presented in Table~\ref{tab:log}; \sfive\ was observed in a total of 84 epochs, and in 11 of them it was monitored intranightly. The duration of all INM sessions but two was about four hours or longer. Some characteristics of the obtained intranight data sets could be found in Table~\ref{tab:inv}, Cols.\,1-6. Zero exposure (bias) and twilight flat-field frames were taken during each observing run.

The primary data reduction consists of bias signal subtraction, flat-field correction, and cosmic ray hits removal. We used the {\sc astropy/ccdproc} data reduction package \citep[][at KAO]{matt_craig_2017_1069648} and locally written {\sc idl} routines (at NAO). The cosmic ray hits removal was performed by means of {\sc l.a.cosmic} \citep{2001PASP..113.1420V,2012ascl.soft07005V}.

\section{Survey data}
\label{sec:asas}

In order to consider our observations in the LTV context, we used the $gV$-band LCs obtained by the All-Sky Automated Survey for Supernovae facility \citep[ASAS-SN;][]{2014ApJ...788...48S,2017PASP..129j4502K}. The ASAS-SN images the entire sky on a nightly basis down to about 17\,mag in the $V$ band. Initially, the ASAS-SN observations were taken in the $V$ band. Then, after some period of overlapping, the observations were carried out in the $g$ band. We used the ASAS-SN LCs from 2012 January 24 to 2023 May 9.

To relate the optical flux changes to the $\gamma$-ray variability of \sfive, we used its three-day and weekly binned $\gamma$-ray LCs in the energy range 0.1--100\,GeV for nearly the same period as the ASAS-SN data. The LCs (with no upper limits included) were retrieved from the {\em Fermi} Large Area Telescope (LAT) Light Curve Repository\footnote{https://fermi.gsfc.nasa.gov/ssc/data/access/lat/LightCurveRepository/} \citep{2023ApJS..265...31A}, which is a database of calibrated LCs of the $\gamma$-ray sources listed in the Fourth {\em Fermi}-LAT catalogue, Data Release 2 \citep{2020arXiv200511208B}. We used a likelihood Test Statistic threshold of four for source detection and a LogParabola spectral type fit with a fixed photon index.

\section{Photometry and variability detection}
\label{sec:photo}

The flux measurements of the blazar and standard stars were performed by means of aperture photometry. Considering the intranight data, we used three different aperture radii, namely 1$\times$, 2$\times$, and 3$\times$FWHM, and chose the one that gave the smallest RMS scatter of the control star LC. In most of the cases, we selected the 2$\times$FWHM radius aperture. Photometry was performed by means of the {\sc astropy/photutils} package \citep[][at KAO]{larry_bradley_2022_6825092} and the {\sc daophot} package \citep[][at NAO]{1987PASP...99..191S}. In particular, we added the world coordinate system information to each frame utilizing the Astrometry.net tool \citep{2010AJ....139.1782L} to detect the objects of interest in the observed field. We selected the 2$\times$FWHM radius aperture for the STV data photometry.

The calibration of the instrumental magnitudes is based on the $BVR$-band standard sequence of \citet{1998A&AS..130..305V}. Regarding the $I$-band magnitudes of the standard stars: star 1 was not calibrated, the magnitude of star 2 was taken from \citet{1997A&A...327...61G}, and the magnitudes of stars 3--8 were taken from \citet{2001AJ....122.2055G}. The $BVRI$-band magnitudes were transformed to the $griz$-band ones using the equations derived by \citet{2005AJ....130..873J} for stars with $R-I < 1.15$\,mag (Table\,\ref{tab:sdss}). We selected stars 3 and 5 as reference ones, and star 6 as a control one. For each frame, we derived the weighted mean zero-point magnitude using the instrumental and catalogue magnitudes of the reference stars. This zero-point was applied in the calibration of the blazar and control star magnitudes.
The outliers (if any) were identified and removed from the final LCs. 

To detect variability in the intranight LCs, we applied the $C$-criterion \citep{1997AJ....114..565J,1999A&AS..135..477R,2010AJ....139.1269D,2017MNRAS.467..340Z,2020MNRAS.498.3013Z}:
\begin{equation}
    C = \frac{\sigma_{\rm blz}}{\sigma_{\rm ctrl}},
\end{equation}
where $\sigma_{\rm blz}$ and $\sigma_{\rm ctrl}$ are the standard deviations of the blazar and control star LCs, respectively.
According to \citet{2017MNRAS.467..340Z}, the target source is variable at 99.5 per cent confidence level or higher if $C \geq 2.576$. To assign a variable/non-variable (Var/NV) status in the case of multi-wavelength data, we generally required the number of bands for which the above condition is fulfilled to be larger/smaller than the number of the bands for which the condition is not fulfilled. In the case of equality the status was set to probably variable (PV).

Application of the $C$-criterion as defined above could result in a false variability detection/non-detection in case of non-negligible difference in brightness between the target source and the control star.
In these regards, a scaling factor accounting for the brightness difference was introduced by \citet{1988AJ.....95..247H}. \citet{2011MNRAS.412.2717J} proposed an alternative definition of the scaling factor, namely:
\begin{equation}
    \kappa^2 = \frac{\left\langle e^2_{\rm blz} \right\rangle}{\left\langle e^2_{\rm ctrl} \right\rangle},
\end{equation}
where $e_{\rm blz}$ and $e_{\rm ctrl}$ are the measurement uncertainties of the blazar and control star instrumental magnitudes, respectively; we used median values instead of mean ones as being more resistant to outliers. We calculated the scaled version of the $C$-criterion as follows:
\begin{equation}
    C_{\kappa} = \left(\frac{1}{\kappa}\right)\frac{\sigma_{\rm blz}}{\sigma_{\rm ctrl}}.
\end{equation}

The percentage variability amplitude, ${\cal A}$, defined in  \citet{1996A&A...305...42H}, was calculated as:
\begin{equation}
    {\cal A} = 100\, \sqrt{~ {(m_\mathrm{max}-m_\mathrm{min}})^2 - 2\langle e^2\rangle_{\rm med} ~} \quad [\rm per~cent],
\end{equation}
where $m_\mathrm{max}$ and $m_\mathrm{min}$ are the blazar maximum and minimum magnitudes, respectively. 

\begin{table}
    \centering
    \caption{Photometric sequence for \sfive\ in the $griz$ bands.}
    \label{tab:sdss}
    \begin{tabular}{ccccc}
    \hline
    Star & $g$ & $r$ & $i$ & $z$ \\
    & [mag] & [mag] & [mag] & [mag] \\
    \hline
    1 & 11.20 & 10.87 &  0.0  &  0.0  \\
    2 & 11.68 & 11.33 & 11.35 & 11.40 \\
    3 & 12.68 & 12.28 & 12.24 & 12.23 \\
    4 & 13.35 & 13.10 & 13.09 & 13.11 \\
    5 & 13.79 & 13.41 & 13.31 & 13.26 \\
    6 & 13.88 & 13.48 & 13.42 & 13.40 \\
    7 & 14.11 & 13.51 & 13.70 & 13.90 \\
    8 & 14.34 & 13.96 & 13.82 & 13.73 \\
    \hline
    \end{tabular}
    \flushleft{{\em Notes.} The magnitude uncertainties are 0.02, 0.03, 0.03, and 0.03\,mag for the $griz$ bands, respectively and represent the RMS residuals of the Johnson-Cousins to SDSS transformations presented in \citet{2005AJ....130..873J}.}
\end{table}

Considering the classification of the $\gamma$-ray activity states, we followed \citet{2014ApJ...789..135W}. According to the authors (i) the upper limit of the {\em quiescent} flux state equals the weighted mean flux, $\langle F \rangle_{\rm wt}$, (ii) the lower limit of the {\em active} state equals $\langle F \rangle_{\rm wt} + \sigma_{\rm wt}$, and (iii) the {\em flaring} state (or the flare) is above $\langle F \rangle_{\rm wt} + 3\sigma_{\rm wt}$ (here $\sigma_{\rm wt}$ is the weighted standard deviation). Between the quiescent and active states the blazar is considered to be in a {\em transitional} one. We also required at least three consecutive data points fulfilling any of these criteria in order to assume that the blazar is in the corresponding state.

\section{Results}
\label{sec:res}

\subsection{Optical light curves combination}
\label{sec:comb}

The LCs of \sfive\ obtained by us in the $V$-band and by ASAS-SN in the $gV$-bands were combined after application of properly derived offsets.
Prior to that, we calculated the nightly weighted-mean blazar magnitude for each night of multiple observations for both sets of data. The larger between the standard uncertainty of the weighted mean and the weighted standard deviation was considered the uncertainty of the nightly mean magnitude. 
The derivation of the offsets was done by comparing the magnitudes of the corresponding LCs for the nights in common and taking the median over the individual nightly offsets.

\begin{figure}
    \includegraphics[width=1.0\columnwidth,clip=true]{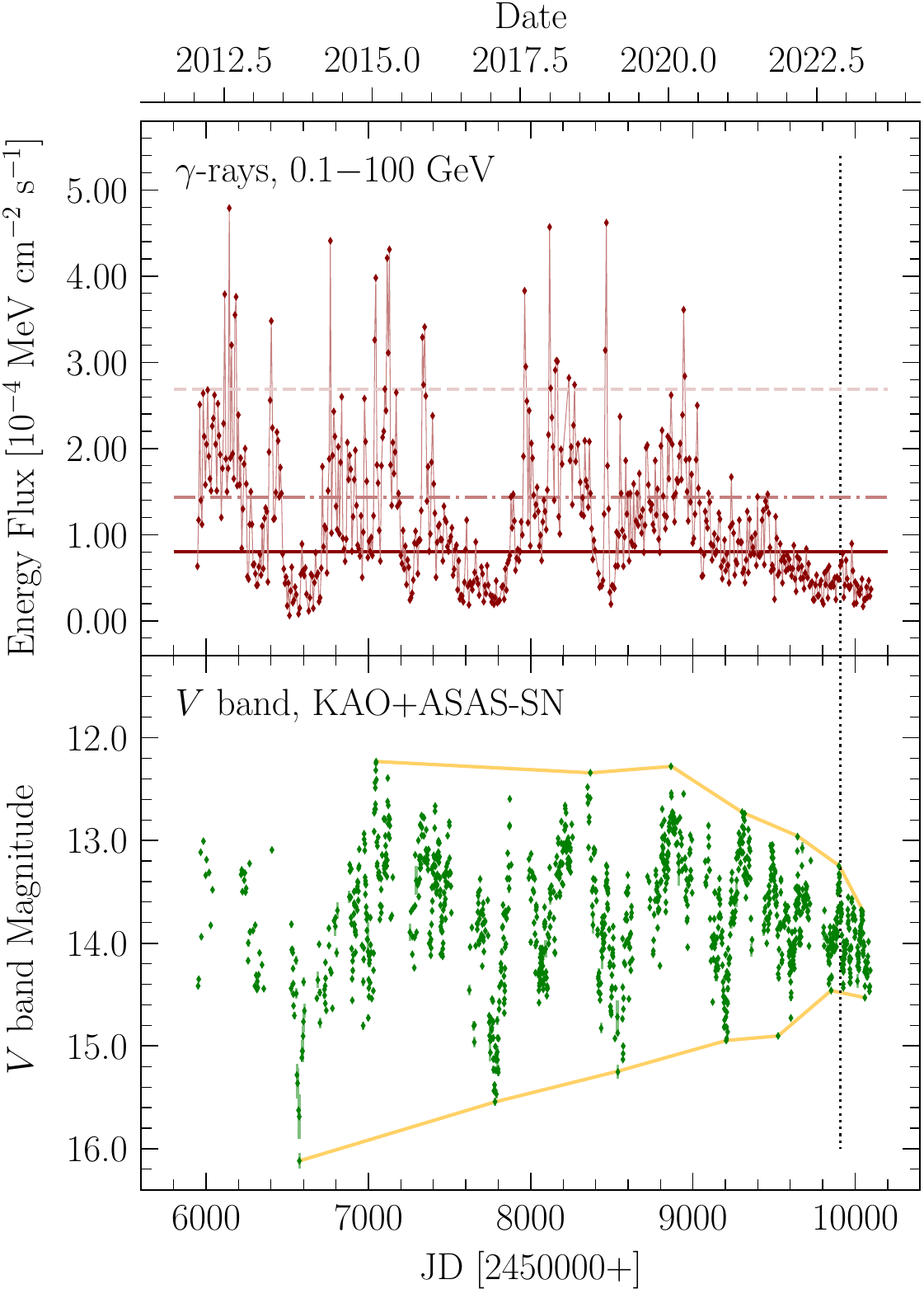}
    \caption{Weekly binned $\gamma$-ray LC of \sfive\ (top panel). The horizontal solid line is the upper limit of the quiescent state, the dash-dotted line is the lower limit of the active state, and the dashed line is the lower limit of the flaring state. Bottom panel: combined long-term $V$-band LC; the orange curves represent the LC envelope. The vertical dotted line in both panels marks the start of our observations.}
\label{fig:lc:ltv}
\end{figure}

\begin{figure}
    \includegraphics[width=1.0\columnwidth,clip=true]{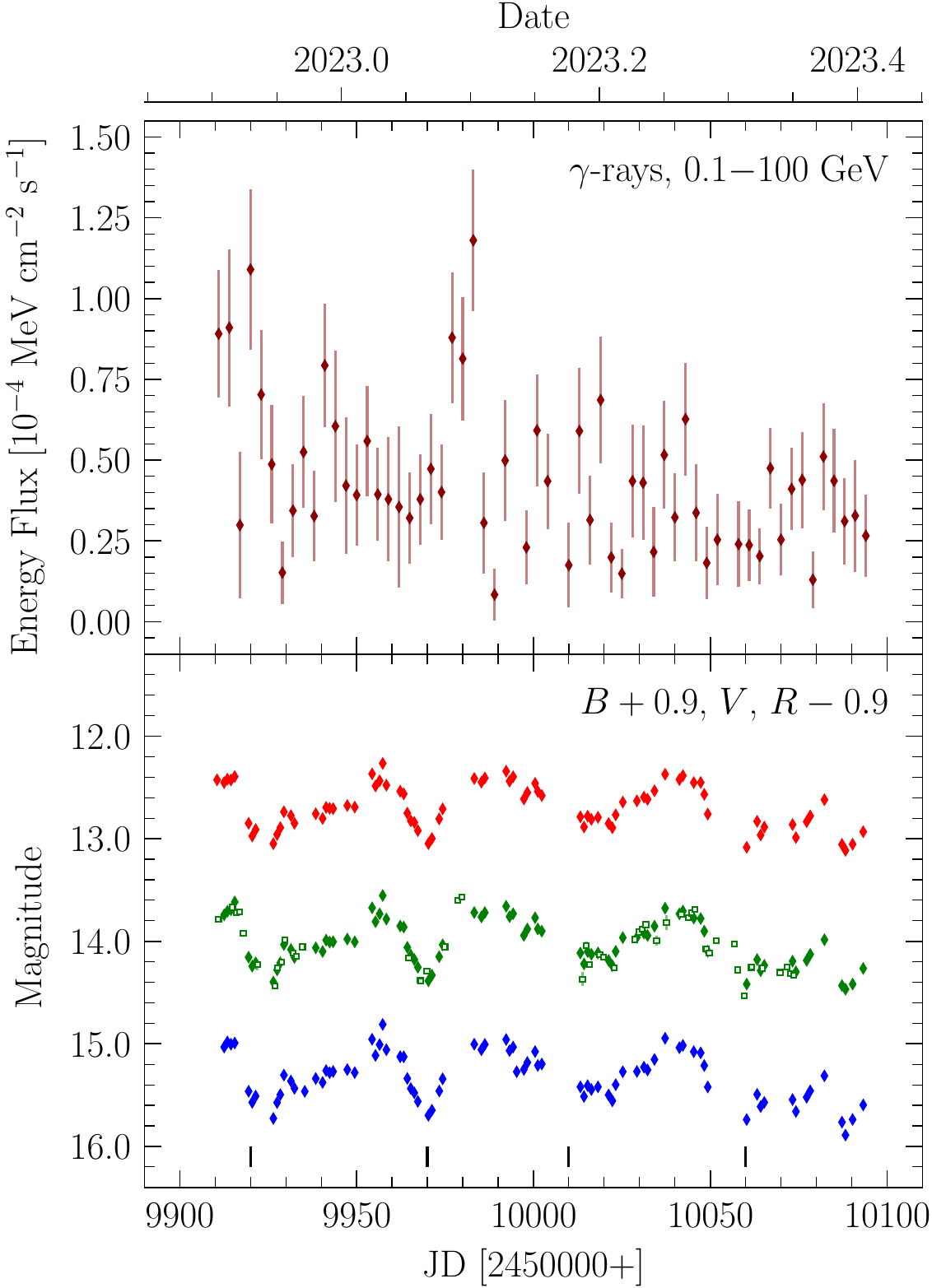}
    \caption{Three-day binned $\gamma$-ray LC of \sfive\ for the period covered by our optical observations (top panel).
    Bottom panel: short-term $BVR$-band LCs (ordered from bottom to top, respectively); the offsets applied for a better presentation are listed. The open squares mark the ASAS-SN data points. The short vertical lines mark the boundaries of the three chunks considered in Sect.~\ref{sec:ccf}.}
    \label{fig:lc:stv}
\end{figure}

\subsection{Long-term variability}
\label{sec:ltv}

The combined long-term $V$-band LC is shown in Fig.~\ref{fig:lc:ltv}. It covers the period from 2012 January 24 to 2023 May 28. We estimated the variability amplitude of the long-term LC to ${\cal A}_V=388.50 \pm 0.08$ per cent. However, during our observing campaign ${\cal A}_V = 97.59 \pm 0.02$ per cent; that is, we registered a decrease of the \sfive\ activity in the optical. The decrease is not sudden, there is a period of a gradual decrease of the variability amplitude since JD $\sim$2459000 (Fig.~\ref{fig:lc:ltv}).

This decrease is more clearly evident on the $\gamma$-ray LC shown in Fig.~\ref{fig:lc:ltv}: prior to JD $\sim$2459000, we estimated the fractional variability\footnote{See \citet{2019Galax...7...62S} for relevant formulae.} of the source to $0.64 \pm 0.03$, while afterwards, we got a fractional variability of $0.49 \pm 0.06$. Following the classification of the $\gamma$-ray flux states, presented at the end of Sect.\,\ref{sec:photo}, \sfive\ is in a quiescent state since JD $\sim$2459600 (Fig.\,\ref{fig:lc:ltv}). The quiescent state lasts for more than $\sim$1.5\,yr and was preceded by a transitional one of nearly the same duration. Thus, according to the long-term LCs our observing campaign took place during a long-lasting period of low activity of \sfive\ in $\gamma$-rays.

Similar quiescent periods of this blazar were observed in the past (around 2013.5--2014.1 and 2016.3--2017.3), but of a shorter duration (about a year or less). The past periods are characterized with sharper boundaries, while now we have a gradual activity decrease. 

\subsection{Short-term variability}
\label{sec:stv}

The short-term $BVR$-band LCs of \sfive\ obtained during our observing campaign are plotted in Fig.~\ref{fig:lc:stv}.
In order to increase the LC time sampling, we added to our $V$-band LC the ASAS-SN $V$-band photometry, transformed as described in Sect.\,\ref{sec:comb}.

We estimated the median time sampling of the $BVR$-band LCs to 1.06\,d. The combined $V$-band LC has a time sampling of 0.99\,d. As mentioned above, the variability amplitude of the combined $V$-band LC during the period of our observing campaign is ${\cal A}_V=97.59 \pm 0.02$ per cent. For the $BR$ bands, we found ${\cal A}_B=107.90 \pm 0.01$ per cent and ${\cal A}_R=84.80 \pm 0.01$ per cent. Thus, the variability amplitude shows decrease from shorter to longer wavelengths.

The optical LCs do not show flaring activity, while the $\gamma$-ray LC shows a flux increase around JD 2459980. The corresponding $V$-band data points are among the brightest ones.

\begin{figure}
    \includegraphics[width=1.0\columnwidth,clip=true]{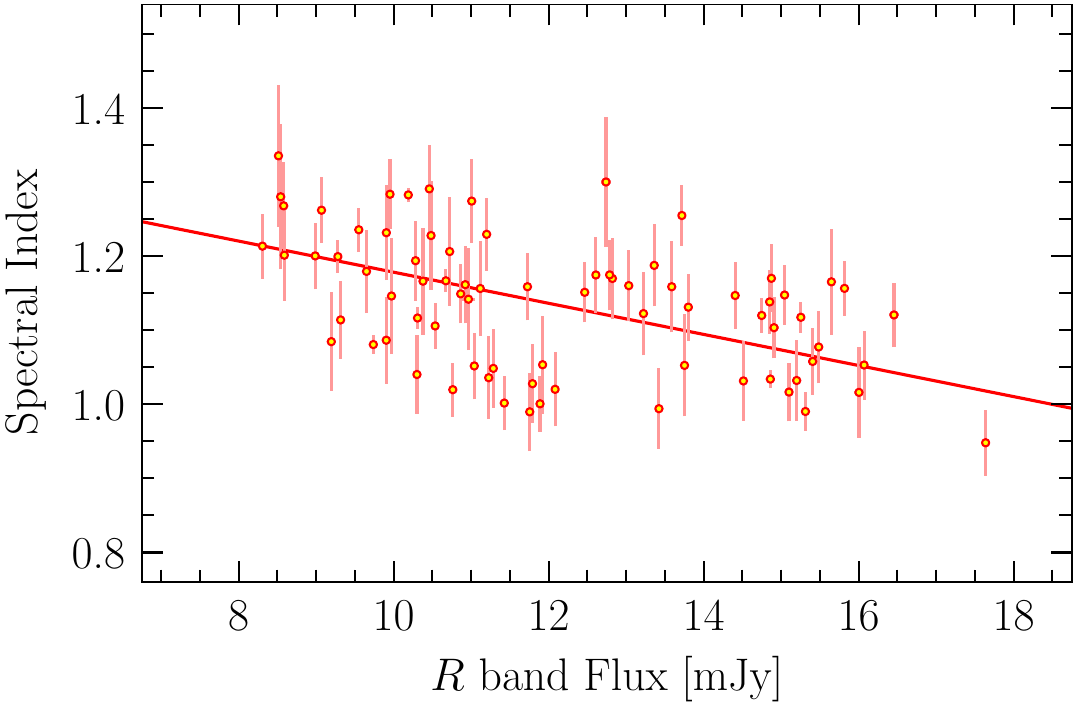}
    \caption{Spectral index $\alpha$ against the $R$-band flux. The solid line is the linear fit to the data.}
    \label{fig:si}
\end{figure}

\begin{table}
    \centering
    \caption{Parameters of the `spectral index~-- flux' relation fitting.} 
    \label{tab:cmd}
    \begin{tabular}{ccc}
    \hline
    Band & Slope        & $r$ \\
         & [mJy$^{-1}$] &    \\
    \hline
    $B$ & $-0.035 \pm 0.005$ & $-0.586$ \\
    $V$ & $-0.027 \pm 0.004$ & $-0.529$ \\
    $R$ & $-0.021 \pm 0.004$ & $-0.470$ \\
    \hline
    \end{tabular}
    \flushleft{{\em Note.} For all fits $p$ is less than $10^{-5}$.}
\end{table}

\subsubsection{Spectral analysis}
\label{sec:cmd}

To study the spectral behaviour of \sfive\ during the period of our observations, we built a SED for each night. The values of the Galactic extinction towards \sfive\ are $\{A_B,A_V,A_R\}=\{0.112,0.085,0.067\}$\,mag. They are calculated using the \citet{2011ApJ...737..103S} recalibration of the \citet{1998ApJ...500..525S} dust infrared emission maps and are taken from the NASA/IPAC Extragalactic Database. The zero-points and effective wavelengths for the Johnson-Cousins system are listed in \citet{1998A&A...333..231B}.

The SEDs were fitted with the linear function
\begin{equation}
\log F_{\nu} = -\alpha \log \nu + \rm const, 
\end{equation}
where $\alpha$ is the spectral index. The dependence of the spectral index on the $R$-band flux is shown in Fig.\,\ref{fig:si}; we can trace a FWB behaviour of \sfive. As a condition for significance, we required the linear Pearson correlation coefficient to be $r \le -0.5$ and the probability to get such a correlation coefficient by a chance to be $p \le 0.01$. The above dependence was fitted by means of an orthogonal distance regression \citep{boggs} in order to take into account the uncertainties of both variables. The fitting results are listed in Table\,\ref{tab:cmd}. We can see that the strength of the anti-correlation is dependent on the band with the strongest anti-correlation for the shortest wavelengths. This is in agreement with the results of \citet{2017A&A...605A..43Y} and \citet{2021RAA....21..102L}. A similar trend of a FWB behaviour of the blazar was reported by \citet[][]{2024PASA...41..103B} and \citet{2025ApJS..277...18L}. Finally, we derived a weighted mean spectral index of $1.139 \pm 0.004$ (a weighted standard deviation of 0.092).
 
\subsubsection{Structure function analysis}
\label{sec:sf}

\begin{figure}
    \includegraphics[width=1.0\columnwidth,clip=true]{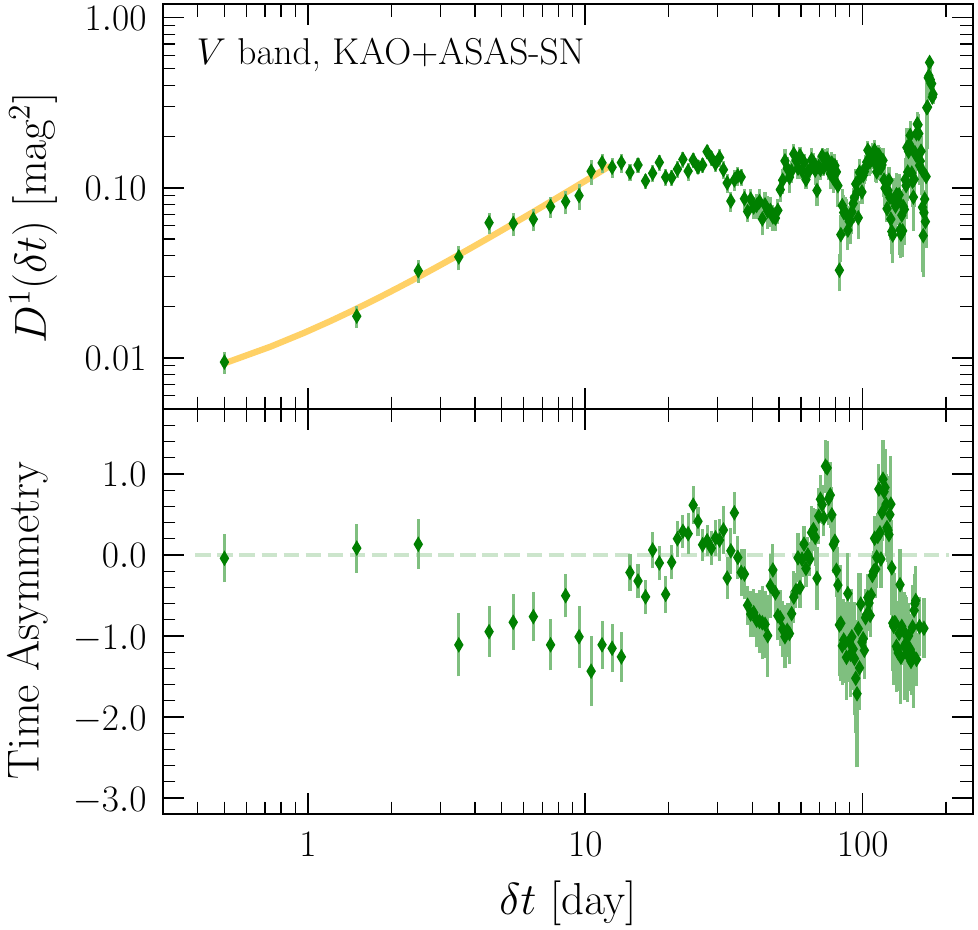}
    \caption{Structure function analysis of the combined $V$-band LC. Top panel: SF with the best fitted SPL model overplotted. Bottom panel: time asymmetry.}
    \label{fig:sf}
\end{figure}

Structure function (SF) is a tool frequently used to analyse unevenly sampled time series \citep{1985ApJ...296...46S}. For a time separation $\delta t$ and bin size ${\rm d}t$, we calculated the first-order SF as:
\begin{equation}
    D^1(\delta t,{\rm d}t) = \frac{1}{N(\delta t,{\rm d}t)} \sum_{j>i}\left[m(t_j)-m(t_i)\right]^2,
\end{equation}
where $N(\delta t,{\rm d}t)$ is the number of pairs $(t_i,t_j)$ for which $\delta t < t_j-t_i < \delta t + {\rm d}t$ and $m$ is the magnitude. The bin size depends on the time sampling of the LC under consideration; we used a bin size of 1\,d. The uncertainties of the SF were calculated as the standard uncertainty of the mean in the bins \citep[see][for discussion about the SF uncertainties]{2020MNRAS.491.5035S}. The value of $\delta t$ in each bin was set to the middle of the bin. The so derived SF is displayed in Fig.~\ref{fig:sf}. If the LC shows quasi-periodic flux variations of period $P$, then there will be SF minima at $\delta t=nP$, where $n=1,2,3$, etc.

We fitted the SF using a single power-law (SPL) model plus a noise term:
\begin{equation}
    D^1(\delta t) = 2\sigma^2_e + D^1_0 \left(\frac{\delta t}{\delta t_0}\right)^{\varrho},
\end{equation}
where $D^1_0$ is the SF value at the fiducial time separation $\delta t_0$ (we arbitrary choose $\delta t_0=1$\,d), $\varrho$ is the SF power-law index (or SF slope), and $\sigma^2_e$ is the variance of the measurement noise. Using this fitting function, we eliminate the influence of the inaccuracies in the observational uncertainty estimation onto the SF slope. This could happen because the noise-free estimate of the SF is $(D^1 - 2\langle e^2\rangle)$ in the case of a Gaussian distribution of measurement uncertainties. The SF fitting is done up to the so called turn-off point at $\delta t=\delta t_{\rm to}$; at this point the SF changes its slope. The SF fit is shown in Fig.~\ref{fig:sf}, and the fitted parameters are listed in Table~\ref{tab:sf}. If the functional dependence of the power spectral density (PSD) on the temporal frequency, $f$, is in the form ${\rm PSD} \propto f^{\,-\varkappa}$, where $\varkappa$ is the PSD power-law index, then the power-law indices are related via $\varkappa - \varrho \simeq 1$. Equality could be obtained under special conditions \citep[e.g.][]{2010MNRAS.404..931E}. On the basis of the SF fits, we obtained $\varkappa\simeq2$ that roughly corresponds to a red-noise type process.

\begin{table}
    \centering
    \caption{Results from the SF analysis.}
    \label{tab:sf}
    \begin{tabular}{cccc}
    \hline
    Band & $\varrho$ & $\delta t_{\rm to}$ & $\langle \rm Time~Asymmetry \rangle_{wt}$ \\
         &           & [d]                 &                                           \\
    \hline
    $B$ & $1.04 \pm 0.14$ & 14.0 & $-0.72 \pm 0.13$ \\
    $V$ & $1.02 \pm 0.11$ & 12.0 & $-0.46 \pm 0.11$ \\
    $R$ & $1.16 \pm 0.14$ & 14.0 & $-0.80 \pm 0.13$ \\
    \hline
    \end{tabular}
    \flushleft{{\em Note.} The weighted mean time asymmetry values were derived using the data for which $\delta t < \delta t_{\rm to}$.}
\end{table}

The SF could be used to study the time asymmetry of LCs as well \citep{1998ApJ...504..671K,2021BlgAJ..34...79B}. In this case, $D^1(\delta t)$ is separated into two parts depending on the sign of $[m(t_j)-m(t_i)]$: the negative differences correspond to the flux rise, while the positive ones correspond to the flux decay. We shall denote the respective SFs as $D^1_{\rm r}$ and $D^1_{\rm d}$.
Given the rise and decay SFs, we can define the time asymmetry as 
\begin{equation}
    {\rm Time\;Asymmetry} \equiv \frac{D^1_{\rm r}-D^1_{\rm d}}{D^1(\delta t)}.
\end{equation}
The time asymmetry of the combined $V$-band LC is shown in Fig.~\ref{fig:sf}; the results for other LCs are similar (see Table~\ref{tab:sf}).
The negative time asymmetry derived by us means that on average the LC shows shallow flux rises and steep flux decays.

\subsubsection{Cross-correlation analysis}
\label{sec:ccf}

To search for time lags, $\tau$, among the optical LCs, we employ the discrete cross-correlation function \citep[DCCF;][]{1988ApJ...333..646E}, which is suitable for unevenly sampled time series. In practice, we used {\sc pydcf}\footnote{https://github.com/astronomerdamo/pydcf}~-- a Python implementation of the DCCF algorithm \citep{2015MNRAS.453.3455R}. DCCF binning was done using a bin width of 1\,d, which is nearly equal to the median time sampling of the $BVR$-band LCs. The time lag itself was estimated as the centroid of the DCCF, that is, by the DCCF-weighting of the lags, for which the DCCF value is larger than or equal to 0.5 times\footnote{Usually, the threshold for centroiding is set to 0.8$\times$$\rm DCCF_{max}$ \citep{2004ApJ...613..682P}, but this choice leaves us with just a few data points to be used in the calculations.} the maximal DCCF value, $\rm DCCF_{max}$. The final time lag and its uncertainty were calculated using the cross-correlation centroid distribution \citep[CCCD;][]{1989MNRAS.236...21M} built on the basis of 2500 runs by means of the flux randomization/random subset selection method \citep{1998PASP..110..660P,2004ApJ...613..682P}. 

\begin{figure}
    \includegraphics[width=1.0\columnwidth,clip=true]{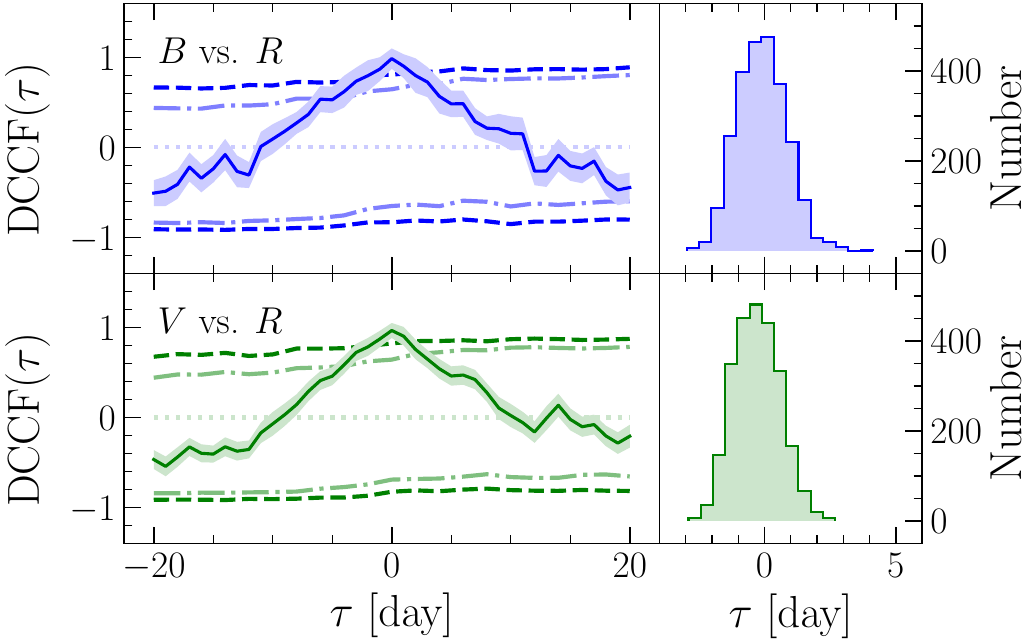}
    \caption{Cross-correlation analysis of the $BVR$-band LCs. The left panel in each plot shows the DCCF and its uncertainties (shaded area), while the right panel shows the corresponding CCCD. The 95 and 99 per cent confidence bands are marked in the left panels with dash-dotted and dashed lines, respectively.}
    \label{fig:dcf}
\end{figure}

To estimate the significance of the DCCF peak, we used the following technique. Firstly, we generated 2500 artificial LCs of the same mean and standard deviation as the observed ones using the method of \citet{1995A&A...300..707T}. The flux distribution of the observed LCs is nearly Gaussian and, so, the usage of that method is justified. If the flux distribution deviates from the Gaussian, then the method of \citet{2013MNRAS.433..907E} for the LC generation is recommended. The slopes of the PSD needed for the LCs generation were taken from the SPL fits (see Sect.\,\ref{sec:sf}). Next, the artificial LCs were interpolated onto the observed JDs to get the same time sampling. The above steps were done for each of the $BVR$ bands. We performed cross-correlation analysis of the generated LCs and built the distribution of the DCCFs. Finally, we calculated the 95th and 99th percentiles of the distribution. The results from the cross-correlation analysis are shown in Fig.~\ref{fig:dcf}. 

\begin{figure}
    \includegraphics[width=1.0\columnwidth,clip=true]{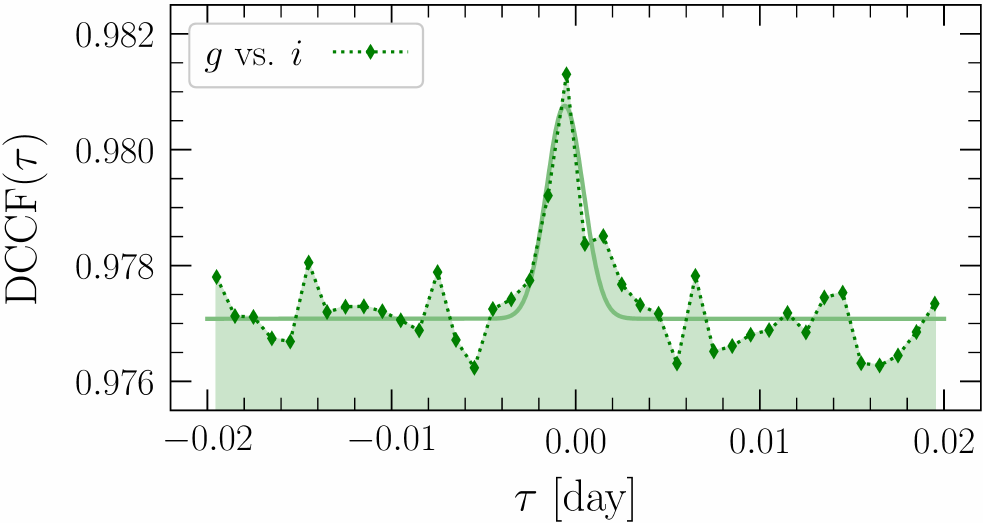}
    \caption{Cross-correlation of the $g$- vs. $i$-band intranight LCs (diamonds). The solid line depicts the Gaussian fitted to the DCCF.}
    \label{fig:dcf:gi}
\end{figure}

We found time lags of $-0.1^{\,+1.1}_{-1.0}$\,d for the $B$- vs. $R$-band DCCF and $0.2^{\,+1.0}_{-1.1}$\,d for the $V$- vs. $R$-band DCCF; that is, there are no lags among the $BVR$-band flux variations within the uncertainties. The peak value of the DCCF is 0.99 for the $B$ vs. $R$ bands and 0.97 for the $V$ vs. $R$ bands, which means the $BVR$-band LCs are strongly correlated.

To refine the time lag estimate, we cross-correlated the intranight $gi$-band LCs, thus, taking advantage of their dense time sampling. The bin width was set to 0.001\,d (1.44\,min) and the resulting DCCF is shown in Fig.~\ref{fig:dcf:gi}. 
The time lag was estimated as the peak position of the Gaussian fitted to the DCCF. 
We obtained $\tau_{gi} = -0.0006 \pm 0.0052$\,d ($-0.9 \pm 7.5$\,min), that is, the lag equals zero to within the quoted uncertainty. Thus, we confirm the lack of lags among the optical LCs during our campaign.

\begin{figure}
    \includegraphics[width=1.0\columnwidth,clip=true]{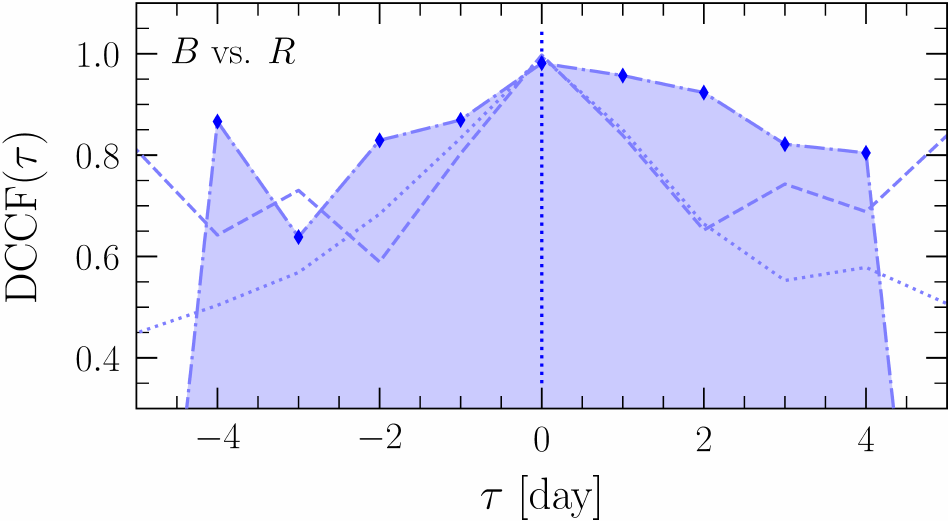}
    \caption{Cross-correlation of the $B$- vs. $R$-band LCs divided into chunks. The dotted, dash-dotted (plus diamonds), and dashed lines are the DCCFs for the chunks marked in Fig.~\ref{fig:lc:stv} from left to right, respectively. The DCCF for the second chunk shows asymmetry towards the positive lags.}
    \label{fig:dcf:chu2}
\end{figure}

\begin{figure}
    \includegraphics[width=1.0\columnwidth,clip=true]{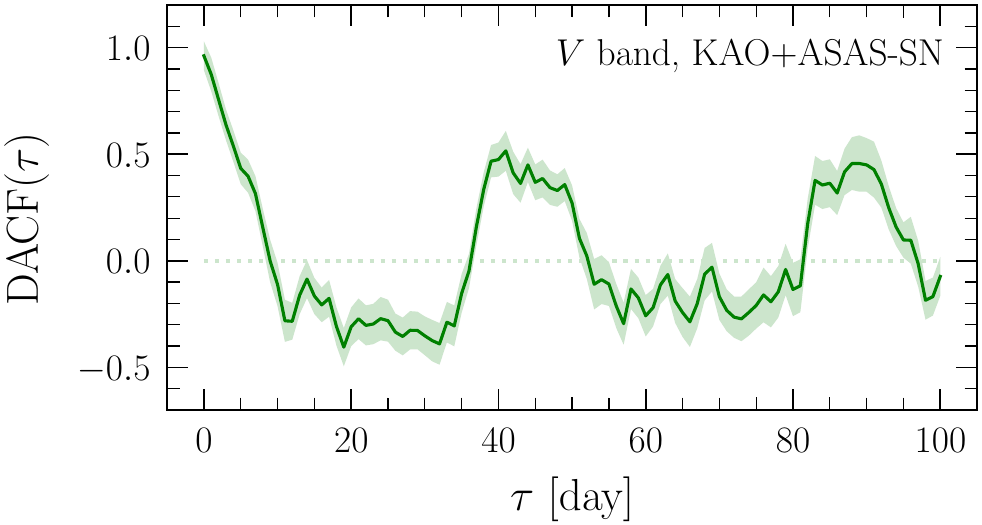}
    \caption{Discrete auto-correlation function for the combined $V$-band LC. The broad peaks for $\tau > 0$ are related to a QPO.}
    \label{fig:acf}
\end{figure}

\begin{table}
    \centering
    \caption{Results from the periodicity analysis.}
    \label{tab:qpo}
    \begin{tabular}{cccc}
    \hline
    Band & Period & Period       & Period \\
         & DACF   & {\sc redfit} & WWZ    \\
         & [d]    & [d]          & [d]    \\
    \hline
    \noalign{\smallskip}
    $B$ & $43.9 \pm 5.7$, $42.5 \pm 5.2$ & $45.5 \pm 5.5$ & $43.0 \pm 6.2$ \\
    $V$ & $43.6 \pm 5.3$, $44.0 \pm 4.9$ & $45.5 \pm 5.1$ & $44.9 \pm 6.9$ \\
    $R$ & $43.9 \pm 5.8$, $42.9 \pm 4.9$ & $45.8 \pm 5.6$ & $43.0 \pm 5.9$ \\
    \noalign{\smallskip}
    Weighted mean & $43.4 \pm 2.1$ & $45.6 \pm 3.1$ & $43.5 \pm 3.6$ \\
    \hline
    \end{tabular}
    \flushleft{{\em Note.} The first and second value of the DACF-estimated period correspond to $n=1,2$, respectively.}
\end{table}

The cross-correlation of the entire LCs results in a time lag, which is a kind of a weight-average over the individual flares (or, more generally, flux peaks) observed on the LCs under consideration \citep{2019ApJ...884...92X,2023ApJS..265...51A}. To avoid this averaging, we divided the LCs into three chunks (see Fig~\ref{fig:lc:stv}) and computed the DCCF for each of the chunks. The DCCF analysis of the LC chunks revealed that the lags are consistent with zero. However, the $B$- vs. $R$-band DCCF of the second chunk, containing the flux peak around JD 2459990, shows asymmetry towards the positive lags (Fig.~\ref{fig:dcf:chu2}); in our notation this means that the $R$-band LC tends to lag the $B$-band one. The $V$- vs. $R$-band DCCF shows similar behaviour, but to a lesser extent. 

We also derived the discrete auto-correlation functions (DACFs) for the $BVR$-band LCs. Figure~\ref{fig:acf} shows the DACF for the combined $V$-band LC; the results for the other bands are similar. The well visible peaks of the DACF for $\tau > 0$ indicate the presence of a QPO~-- the peaks correspond to $\tau = nP$, where $n=1,2$. The centroid values of the lags corresponding to the DACF peaks were determined in the same way as for the DCCF. The DACF-estimated periods are listed in Table~\ref{tab:qpo}. The period uncertainties are the half-width at the half-maximum of the fitted Gaussians to the peaks \citep[e.g.][]{2020MNRAS.492.5524O}; the possible asymmetry of the fitted peaks was ignored. We also determined the zero-crossing time-scales for the $BVR$-band DACFs~-- 11.3, 9.0, and 11.2\,d, respectively; this time-scale equals the time lag at which the DACF crosses the zero-correlation line for the first time.

\begin{figure}
    \includegraphics[width=1.0\columnwidth,clip=true]{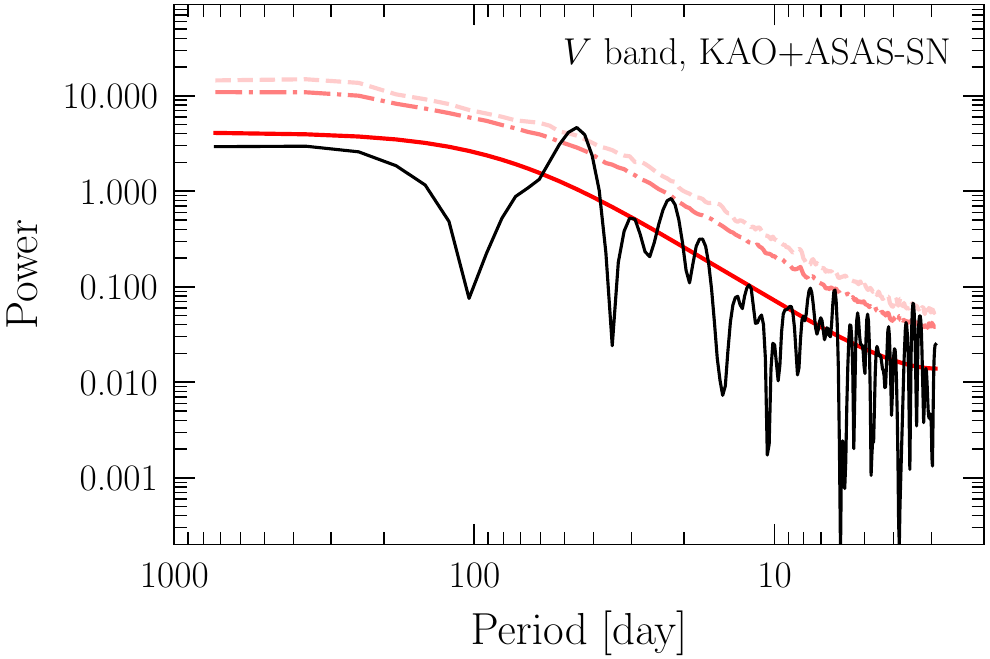}
    \caption{Periodicity analysis of the combined $V$-band LC based on the {\sc redfit} programme. We plot the bias-corrected spectrum of the LC (black line), theoretical red-noise spectrum (solid red line), as well as 95 and 99 per cent significance levels established by means of Monte Carlo simulations (dash-dotted and dashed lines, respectively).}
    \label{fig:redfit}
\end{figure}

\begin{figure}
    \includegraphics[width=1.0\columnwidth,clip=true]{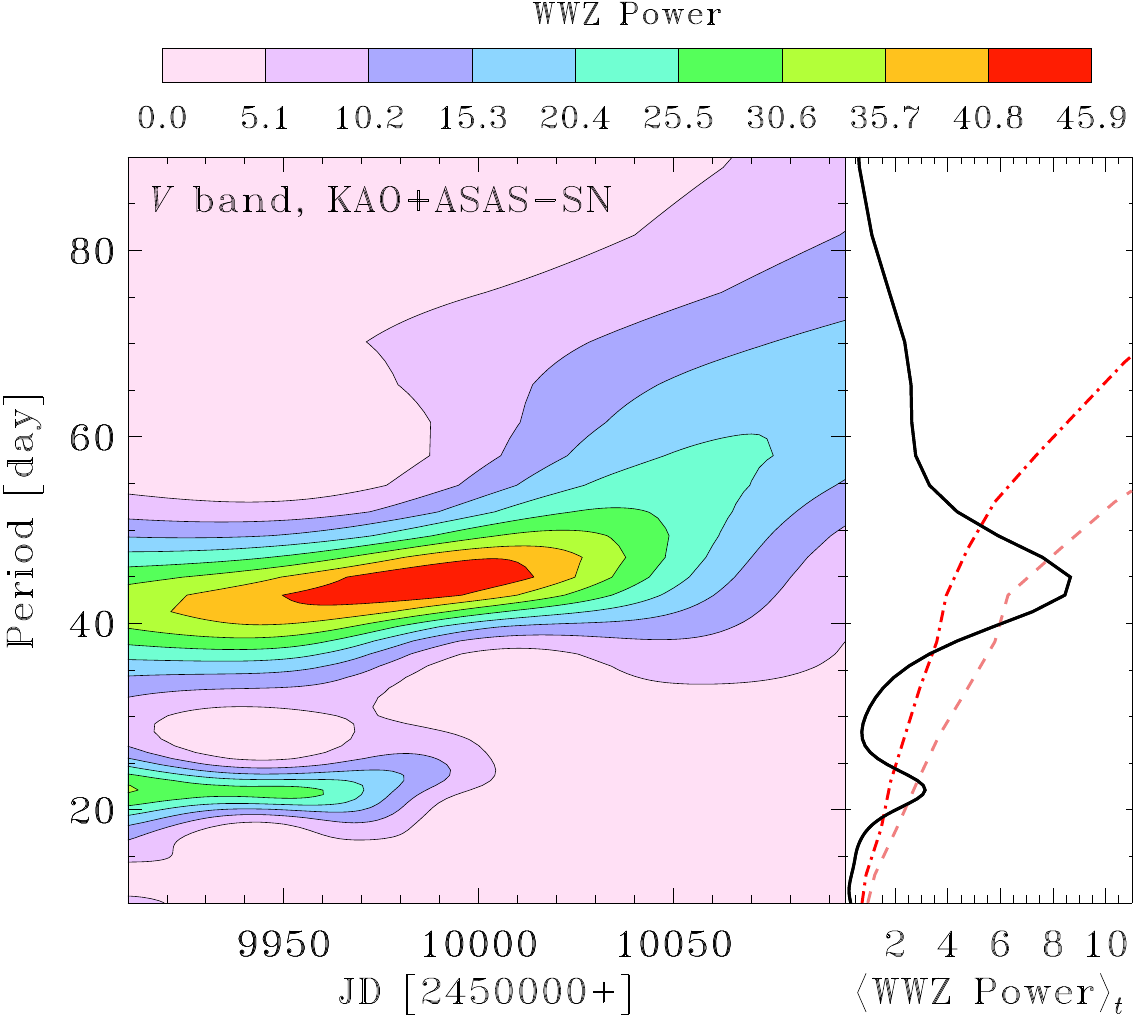}
    \caption{Periodicity analysis of the combined $V$-band LC using the WWZ method. Left panel: the WWZ power as a function of the period and observing time. Right panel: the time-averaged WWZ power against the period (black line). The significance levels of 95 and 99 per cent are plotted with dash-dotted and dashed red lines, respectively.}
    \label{fig:wwz}
\end{figure}

\begin{figure}
    \includegraphics[width=1.0\columnwidth,clip=true]{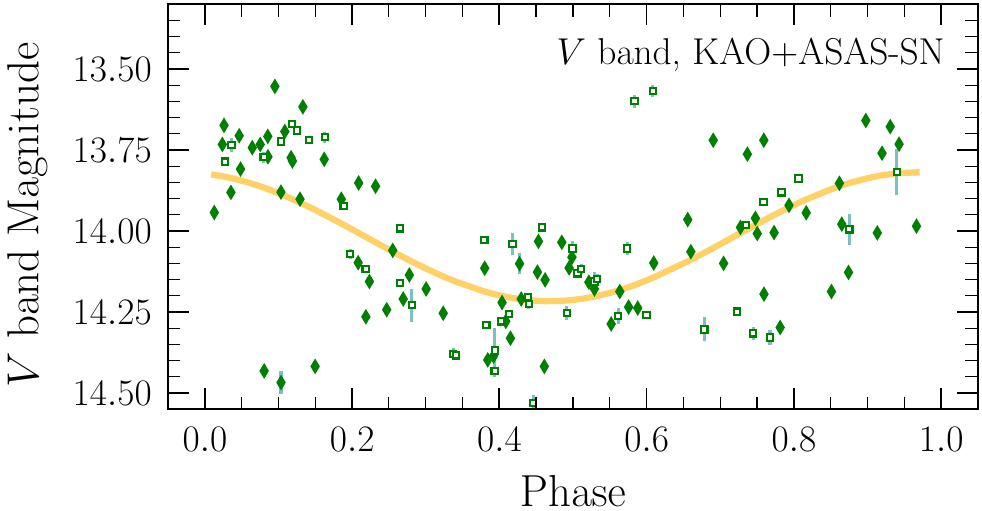}
    \caption{The combined $V$-band LC folded with a period of 43.5\,d; the symbols are the same as in Fig.\,\ref{fig:lc:stv}. The solid curve represents a sine wave of the same period fitted to the LC.}
    \label{fig:folded}
\end{figure}

\subsubsection{Periodicity analysis}
\label{sec:qpo}

We applied the {\sc redfit} programme \citep{2002CG.....28..421S} and the weighted wavelet Z-transform technique \citep[WWZ;][]{1996AJ....112.1709F,1998BAMS...79...61T} to study further the QPO.

The {\sc redfit} programme could be used for the search of periodicity in unevenly sampled LCs by means of spectral-peak testing against the red-noise background. The red-noise background spectrum is estimated using a first-order autoregressive process (AR1) fitted to the LC. Before the calculations, we ran non-parametric tests and found that the AR1 model is appropriate description of the LCs; in addition, the LC data points are not too clustered in time. These two conditions should be fulfilled in order {\sc redfit} to be run properly \citep[see][for details]{2002CG.....28..421S}. We used a port of the original {\sc redfit} programme, which is included in the dendrochronology programme library (dplR\footnote{https://rdrr.io/cran/dplR/}) within the {\sc r} statistical environment \citep{2008Dendr..26..115B}.

We ran {\sc redfit} using a Welch spectral window, oversampling factor of 4, and 2500 Monte Carlo simulations to estimate the peak significance. The outcome for the combined $V$-band LC is shown in Fig.~\ref{fig:redfit}; the results for the other bands are similar (see Table~\ref{tab:qpo}). As a further result of the {\sc redfit} run, we got an estimate of the decorrelation time-scale, which characterizes the AR1 process \citep{2002CG.....28...69M,2015MNRAS.451.4328K}. This time-scale measures the time interval after which the contribution of the previous LC data points could be considered negligible, that is, it is a measure of the process `memory'. The values we obtained are 12.1, 12.4, and 10.9\,days for the $BVR$-band LCs, respectively.

\begin{figure*}
    \includegraphics[width=0.341\textwidth,clip=true]{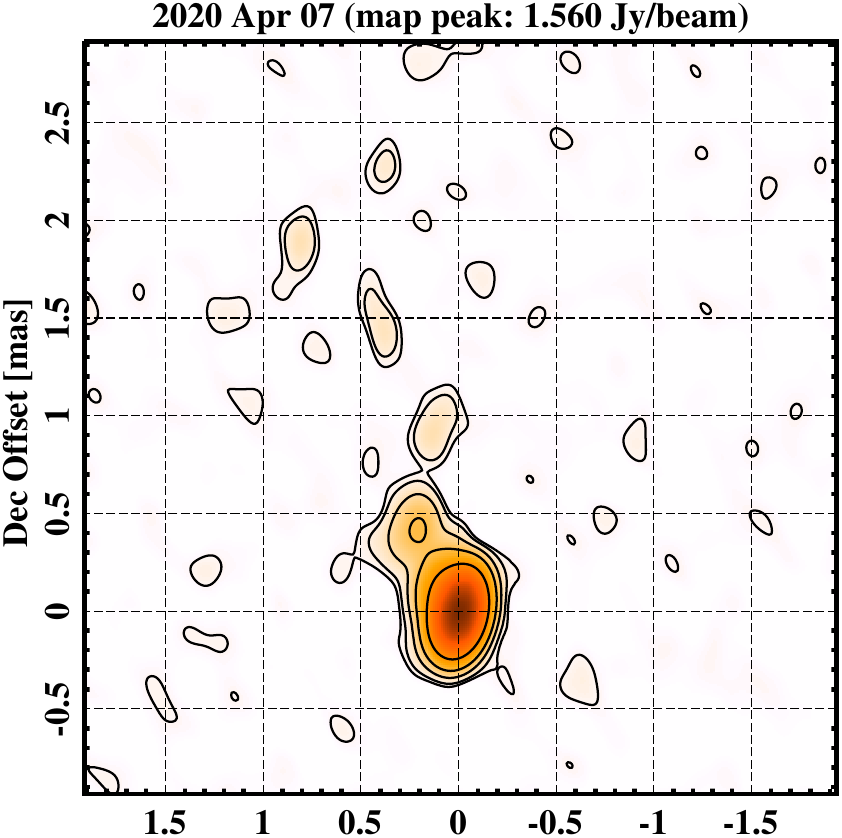}\hspace{0.05em}
    \includegraphics[width=0.324\textwidth,clip=true]{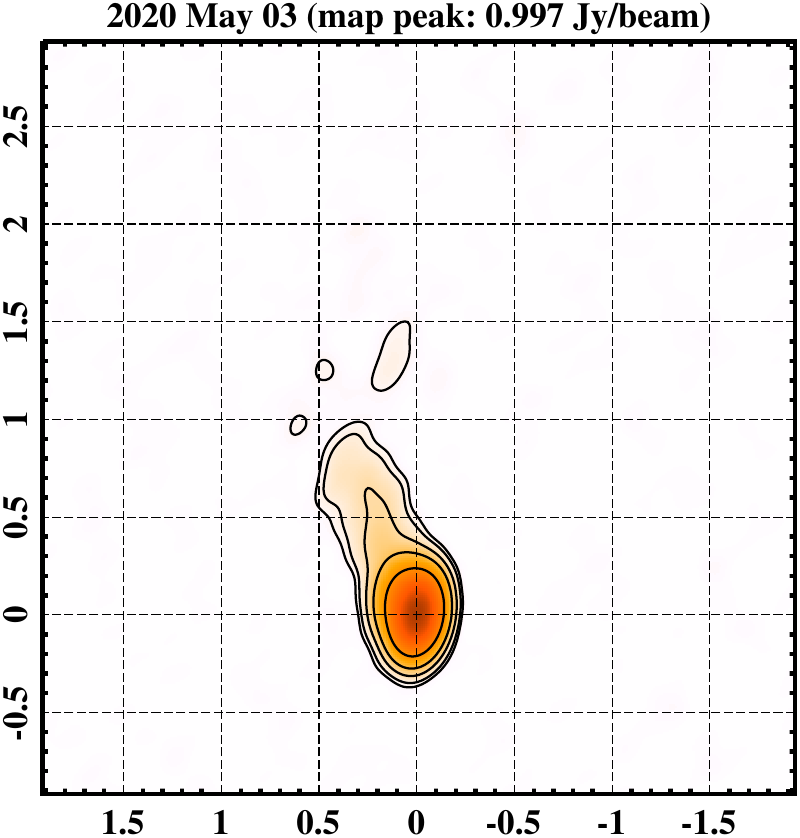}\hspace{0.05em}
    \includegraphics[width=0.324\textwidth,clip=true]{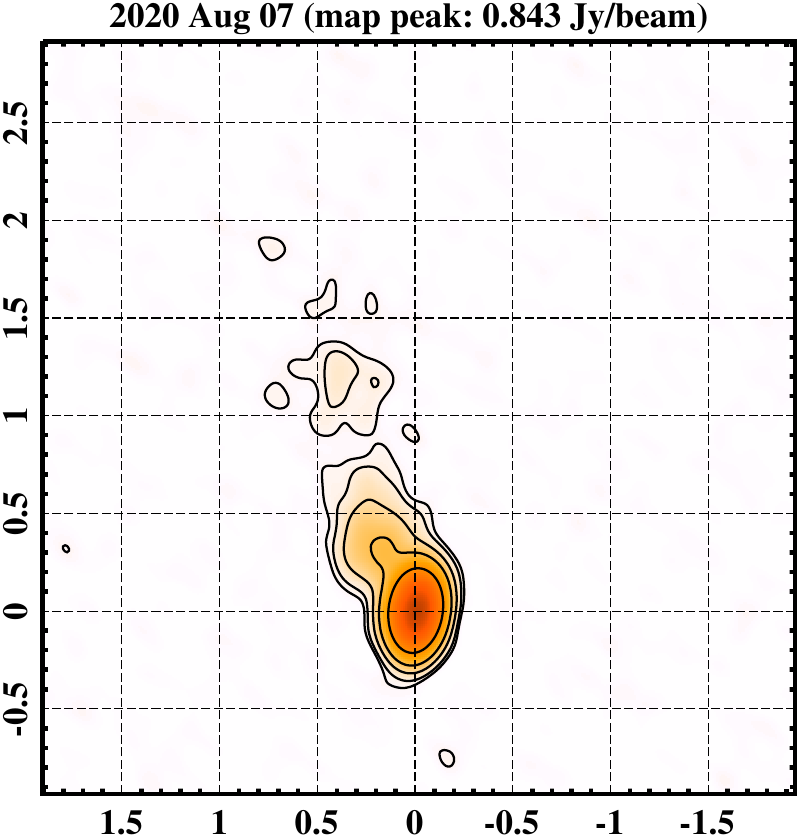}\vspace{0.25cm}
    \includegraphics[width=0.341\textwidth,clip=true]{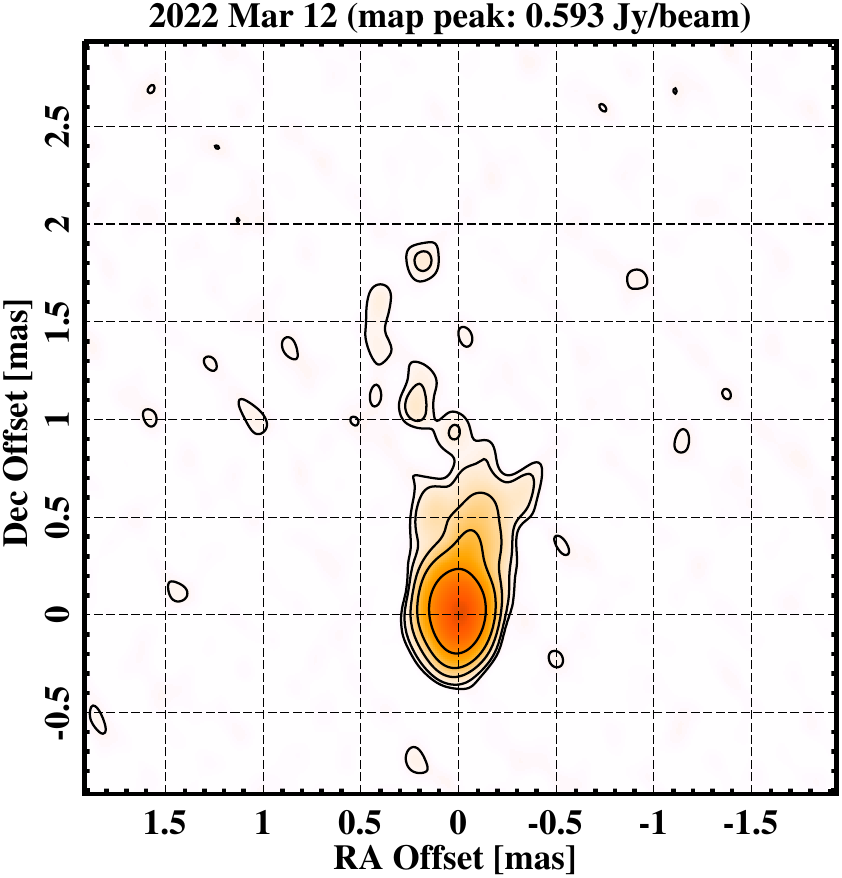}\hspace{0.05em}
    \includegraphics[width=0.324\textwidth,clip=true]{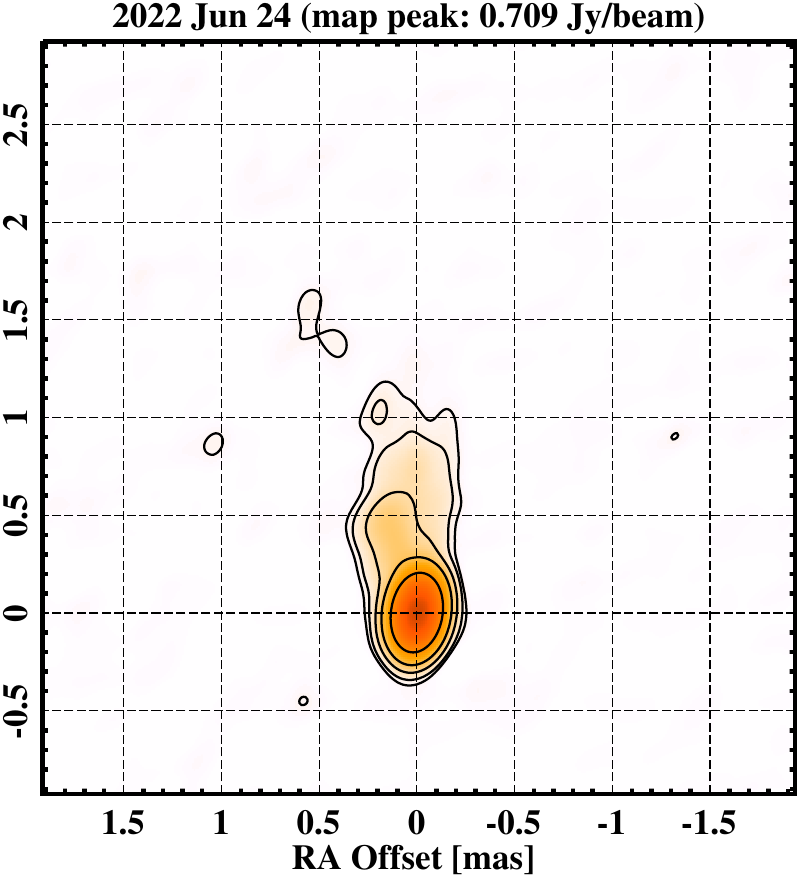}\hspace{0.05em}
    \includegraphics[width=0.324\textwidth,clip=true]{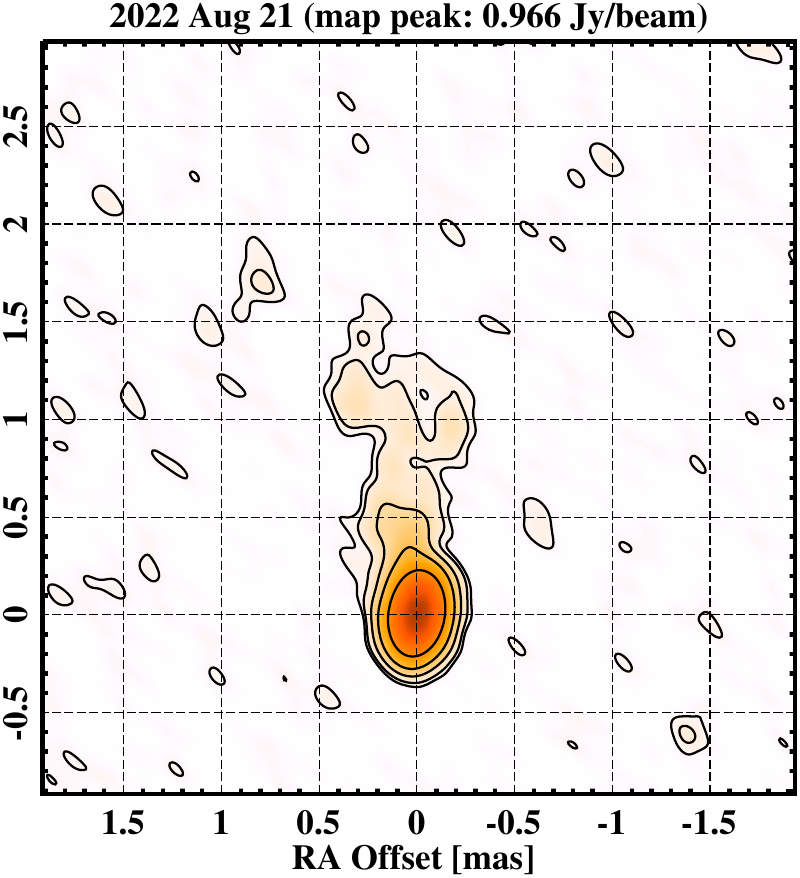}
    \caption{Contour maps of the \sfive\ radio images taken at six epochs with the Very Long Baseline Array at 43\,GHz (the project BEAM-ME); the images were downloaded from https://www.bu.edu/blazars/BEAM-ME.html. The contour levels are 1.25, 2.5, 7.5, 22.5, and 100\,mJy\,beam$^{-1}$. These sequences trace the ejection of new components during 2020 April and 2022 March and their evolution (top and bottom panel rows, respectively).}
    \label{fig:vlba}
\end{figure*}

We employ WWZ technique, based on the Morlet wavelet, that is suitable for unevenly sampled LCs. In addition, the WWZ calculates the power as a function of the observing time and period, that is, it can detect transient QPOs. This makes it a powerful tool for the QPO evolution study. We used a Python code\footnote{https://github.com/eaydin/WWZ} \citep{m_emre_aydin_2017_375648} to perform WWZ; the value of the constant, which determines how fast the wavelet decays, was fixed to 0.0125 \citep{1996AJ....112.1709F}. To estimate the significance of the time-averaged WWZ peak(s), we used the same technique as in Sect.~\ref{sec:ccf}. The WWZ analysis of the combined $V$-band LC is shown in Fig.~\ref{fig:wwz}; the results for the other LCs are similar (see Table~\ref{tab:qpo}). 
Thus, the WWZ analysis points to the transient nature of the QPO detected in the flux variations of \sfive\ during our observing campaign. 

For each technique used to estimate the QPO period, we list the weighted mean value over the bands in Table~\ref{tab:qpo}. The periods obtained using different techniques are equal to within the uncertainties. Taking a median over the weighted mean results, we got a final period of $43.5 \pm 3.6$\,d.
In Fig.~\ref{fig:folded}, we show the combined $V$-band LC folded with the median period. 

\subsection{Intranight variability}
\label{sec:inv}

The blazar \sfive\ was monitored intranightly for a total of 11 nights with a median duration of 4.7\,h per night. The MWL data at KAO were taken quasi-simultaneously (a repeated sequence of $gi$- or $griz$-band frames), while the NAO data are simultaneous. The intranight LCs are presented in Fig.~\ref{fig:lc:inv} (Appendix\,\ref{app:lc}). Generally, the LCs show quite smooth flux behaviour without indications for even low-amplitude flares. The $C$-criterion was applied to the LCs in order to estimate the presence of INV (see Sect.~\ref{sec:photo}). The results are summarized in Table~\ref{tab:inv} along with some other characteristics of the intranight LCs. The calculated DCs of \sfive\ are as follows: 10.3 per cent if we consider the PV case non-variable and 20.5 per cent if the PV case is regarded variable. The DC itself is calculated as the total INV duration over the total INM duration \citep{1999A&AS..135..477R}. 

We note that the DC values may vary depending on the variability detection technique used and on the way of DC estimation. The conservative tests, like $C$-criterion, would reject some variability detections found by other, less conservative tests, like $\chi^2$-test \citep[see][for an example in this context]{2015ApJS..218...18D}. The most popular way of estimating DC is the one proposed by \citet{1999A&AS..135..477R}, based on weights, equal to the inverse of the INM duration.
This technique, however, was found to be biased \citep{2021Galax...9..114W}. Another approach uses the number of nights of INV over the number of nights of INM (applicable if no information about the INM duration is available). 

The observed intranight LCs show large variety of shapes that reflect the complex and stochastic nature of the blazar variability. The variability detection tests, like $C$-criterion, $F$-test, etc., are not sensitive to the LC shapes, they just detect whether the flux changes satisfy the predefined criteria and assign to the corresponding night variable or non-variable blazar status. In case of variability, two LC patterns could be discerned on an intranight scale: flaring and smooth. The LCs with variability status Var+PV obtained in this work could be considered of a smooth type.
Therefore, if we present the INV DC as consisting of a flaring and smooth component, the first one is zero and the second one equals the DC value derived above.
Detailed discussion on the LC shapes will be presented elsewhere.

For the night with detected INV, we have single-band data and, hence, we cannot study the colour behaviour of the blazar. Considering the PV case, we built the colour-magnitude diagram $(g-i)$ vs. $(g+i)/2$ and got an anti-correlation that is not significant ($r=-0.10$, $p=0.17$).

\section{Discussion}
\label{sec:disc}

In this paper we have presented the results from the MWL monitoring of the blazar \sfive\ for the period 2022 November 26~-- 2023 May 28 on short-term and intranight time scales using optical telescopes in Egypt and Bulgaria. We collected MWL data in a total of 84 epochs, and for 11 of them we performed INM. We appended our data with the properly transformed ASAS-SN survey $gV$-band photometry to study the \sfive\ flux variations on long-term time-scales and to increase the time sampling of our short-term $V$-band data. In addition, {\em Fermi}-LAT $\gamma$-ray data were used on long- and short-term time scales.

\begin{figure}
    \includegraphics[width=1.0\columnwidth,clip=true]{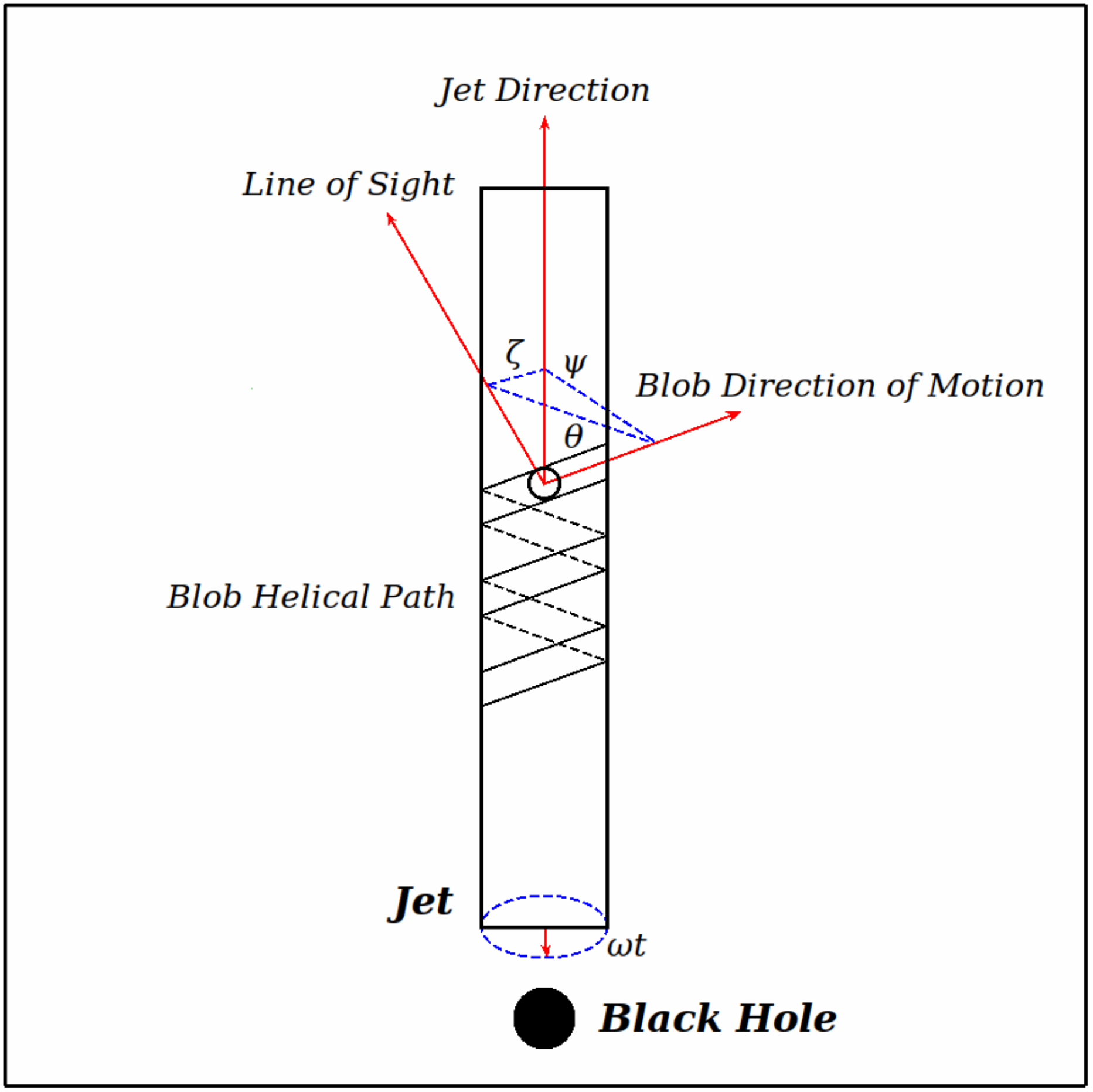}
    \caption{Sketch of the blob helical motion within the blazar jet. The directions used to define the relevant angles are denoted (see text for details).}
    \label{fig:helix}
\end{figure}

\subsection{Long-term variability}

The main feature of the long-term optical and $\gamma$-ray LCs of \sfive\ is the gradual activity decrease (better traced in the $\gamma$-rays) started at JD $\sim$2459000. In the context of the LTV, our observing campaign took place during a low-activity state of \sfive\ in $\gamma$-rays. Generally, low-activity states provide an opportunity for studying the baseline jet emission without a significant contamination by flaring components \citep[e.g.][]{2023ApJS..266...37A,2024PASA...41..103B}.
Similar long-lasting, low-activity states were observed in other blazars, as well: Mrk\,501 \citep{2023ApJS..266...37A}, PKS\,1510--089 \citep{2010ApJ...721.1425A}.

During the previous low-state period of \sfive\ around 2016.3--2017.3, two superluminal knots were ejected \citep[B11 and B12;][]{2022ApJ...925...64K}. To check the mm-wavelength behaviour of \sfive\ over the present period of low activity, we visually inspected the Very Long Baseline Array 43\,GHz data\footnote{https://www.bu.edu/blazars/BEAM-ME.html}, collected by the Boston University Blazar Group \citep{2016Galax...4...47J,2017ApJ...846...98J,2022ApJS..260...12W}. We identified two episodes of new jet component ejection: during 2020 April and 2022 March (Fig.\,\ref{fig:vlba}).
Therefore, again we observed ejection events in \sfive\ without a significant optical/$\gamma$-ray flux enhancement. According to \citet{2016Galax...4...47J} while the bulk of flares occur near the time of knot ejection, more than half of ejection events are not responsible for $\gamma$-ray flares.

Using the long-term ASAS-SN and {\em Fermi}-LAT LCs of \sfive, \citet{2023MNRAS.519.6349D} found that the optical flares lead the $\gamma$-ray ones by $5.89 \pm 33.53$\,d. Given the quoted uncertainty, we can assume that the corresponding emitting regions are co-spatial.

\subsection{Short-term variability}

On short-term time-scales, we observed moderately FWB chromatic flux variations with variability amplitude of $97.59 \pm 0.02$ per cent in the $V$ band.
The LCs are strongly correlated with no time lags among the individual bands.

\begin{figure*}
    \includegraphics[width=0.54\textwidth,clip=true]{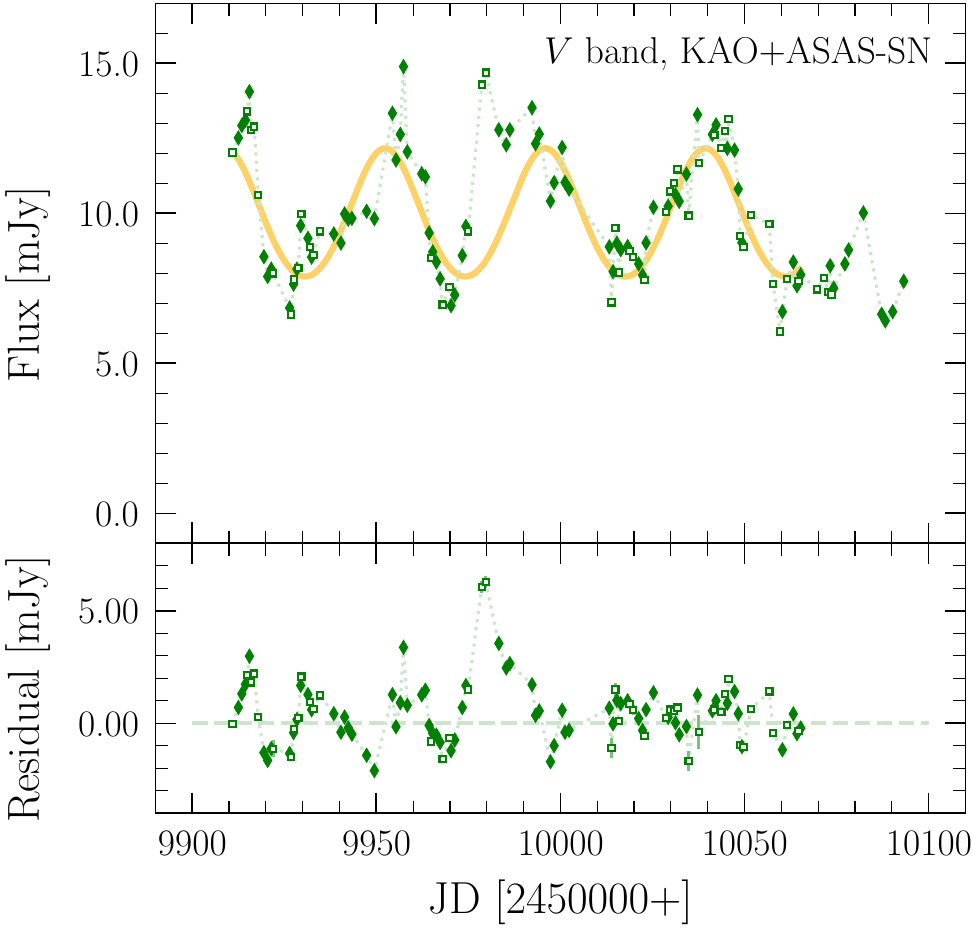}
    \includegraphics[width=0.456\textwidth,clip=true]{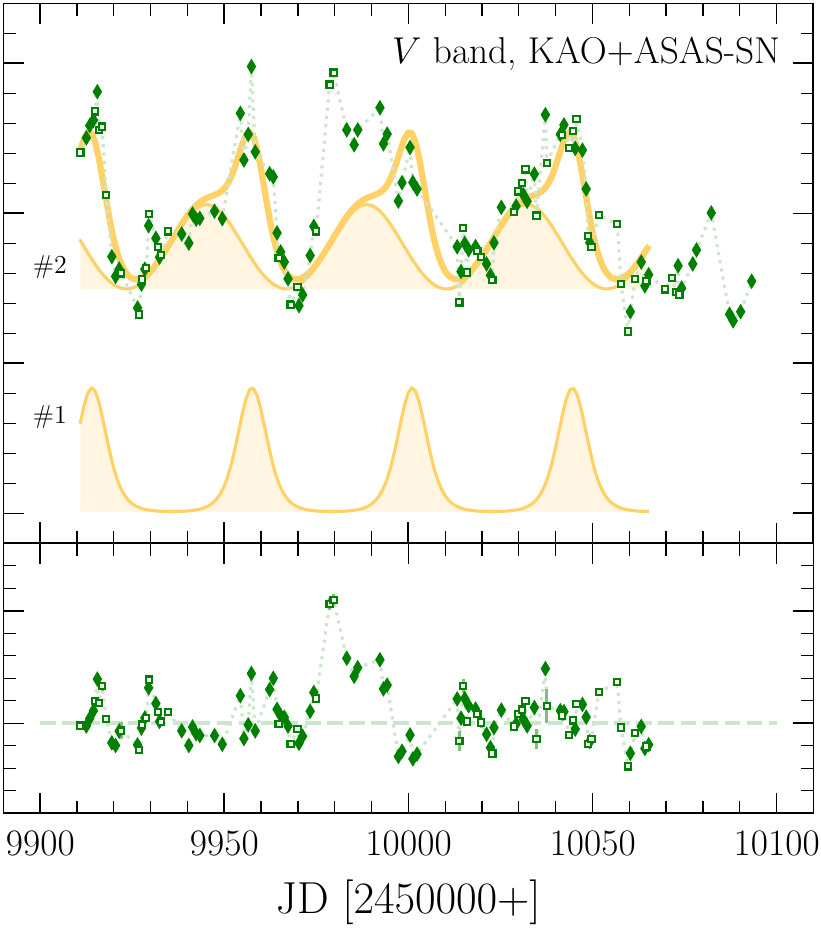}
    \caption{Results from the modelling of the combined $V$-band LC (top panels, the symbols are the same as in Fig.\,\ref{fig:lc:stv}). The orange curves are the model LCs produced by a single blob (left panel) or by a couple of blobs (right panel). The shaded curves in the top right panel depict the model LCs produced by the blobs \#1 and \#2. The fitting residuals are presented in the bottom panels.}
    \label{fig:helix:fit}
\end{figure*}

\subsubsection{Periodicity interpretation}
\label{sec:qpo:d}

The periodicity analysis of the data resulted in a detection of a transient QPO with a period of $43.5 \pm 3.6$\,d. The value we found is equal within the uncertainties to the one derived by \citet{2024ApJ...961..180L}, $44\pm6$\,d, on the basis of the $gri$-band data of \citet[][]{2020ApJS..247...49X} taken from 2017 November to 2018 May.

For the interpretation of the observed quasi-periodicity, we shall assume a model of a blob helical motion (Fig.\,\ref{fig:helix}). The blob viewing angle, $\theta$, depends on time, $t$, as follows \citep[e.g.][]{2018NatCo...9.4599Z}:
\begin{equation}
    \cos\theta(t) = \cos\zeta\cos\psi + \sin\zeta\sin\psi\cos\left(\frac{2 \pi t}{P} + \varphi_0\right),
\end{equation}
where $\zeta$ is the inclination angle, $\psi$ is the pitch angle, and $\varphi_0$ is the azimuth angle at $t=0$.
The Doppler factor also depends on time:
\begin{equation}
    \delta(t) = \frac{1}{\Gamma\left[1-\beta\,\cos\theta(t)\right]},
\end{equation}
where $\beta$ is the blob velocity in units of speed of light, $c$, and the bulk Lorentz factor $\Gamma$ equals:
\begin{equation}
    \Gamma = \frac{1}{\sqrt{~ 1-\beta^2 ~}}.
\end{equation}
The blazar flux in the optical could be expressed as:
\begin{equation}
    F'(\nu')=F'(\nu'_0)\left(\frac{\nu'}{\nu'_0}\right)^{-\alpha},
\end{equation}
where $F'(\nu'_0)$ is the flux at the fiducial frequency $\nu'_0$.
The observed flux in the presence of Doppler boosting is:
\begin{equation}
    F(t,\nu)=F'(\nu')\,\delta(t)^{\,\alpha+\varsigma},
\end{equation}
where the Doppler boost exponent is $\varsigma=2$ for a smooth jet flow and $\varsigma=3$ for a resolved blob within the jet \citep[e.g.][]{2007ApJ...658..232C}. 
Finally, to get the flux in a predefined band $\{\nu_1,\nu_2\}$, we need to integrate $F(t,\nu)$ over this band taking into account the filter transmission curve \citep[e.g.][]{2016ApJ...833...98C}.

The combined $V$-band LC was modelled based on the above formalism.
We set $\Gamma = 14.0$ \citep[a weighted mean over the values reported in][]{2017ApJ...846...98J}, $\alpha = 1.139$ and $P = 43.5$\,d (as derived in this work), as well as $\varsigma = 3$.
The resulting fit is shown in Fig.~\ref{fig:helix:fit}, left panel  (the fitting interval was set according to the WWZ periodicity analysis). The best-fitting parameters are: $\zeta \simeq 1\fdg6$, $\psi \simeq 0\fdg3$, and $\varphi_0 \simeq 75\fdg4$. The calculated Doppler factor varies from 23.4 to 25.4 with a mean of 24.3.

Given the (negative) time asymmetry of the LCs, we considered a two-blob model as well.
The above formulae were applied to each blob individually, and the total flux was obtained as $F(t)=F_1(t)+F_2(t)$, where the indices stand for the blobs \#1 and \#2. The resulting fit is shown in Fig.~\ref{fig:helix:fit}, right panel, and the fitted model parameters are listed in Table~\ref{tab:helix:fit}. We assume that a change in a model is acceptable if the Bayesian information criterion decreases by 10 or more \citep[e.g.][]{2023ApJS..265...51A}. The Bayesian information criterion dropped from 659 to 632 after the inclusion of the second blob, so, its addition is justified. The Doppler factor varies from 12.4 to 27.9 with a mean of 18.8 for blob \#1 and from 21.1 to 22.4 with a mean of 21.7 for blob \#2. Using two moving blobs, we modelled successfully the observed LC with an exception of the region around JD 2459980, where the residuals are systematically positive (see Fig.\,\ref{fig:helix:fit}). This indicates a deviation of the flux variations with respect to the pure geometric model adopted by us.

\subsubsection{The synchrotron flare and the emission region parameters}
\label{sec:syn}

\begin{figure}
    \includegraphics[width=1.0\columnwidth,clip=true]{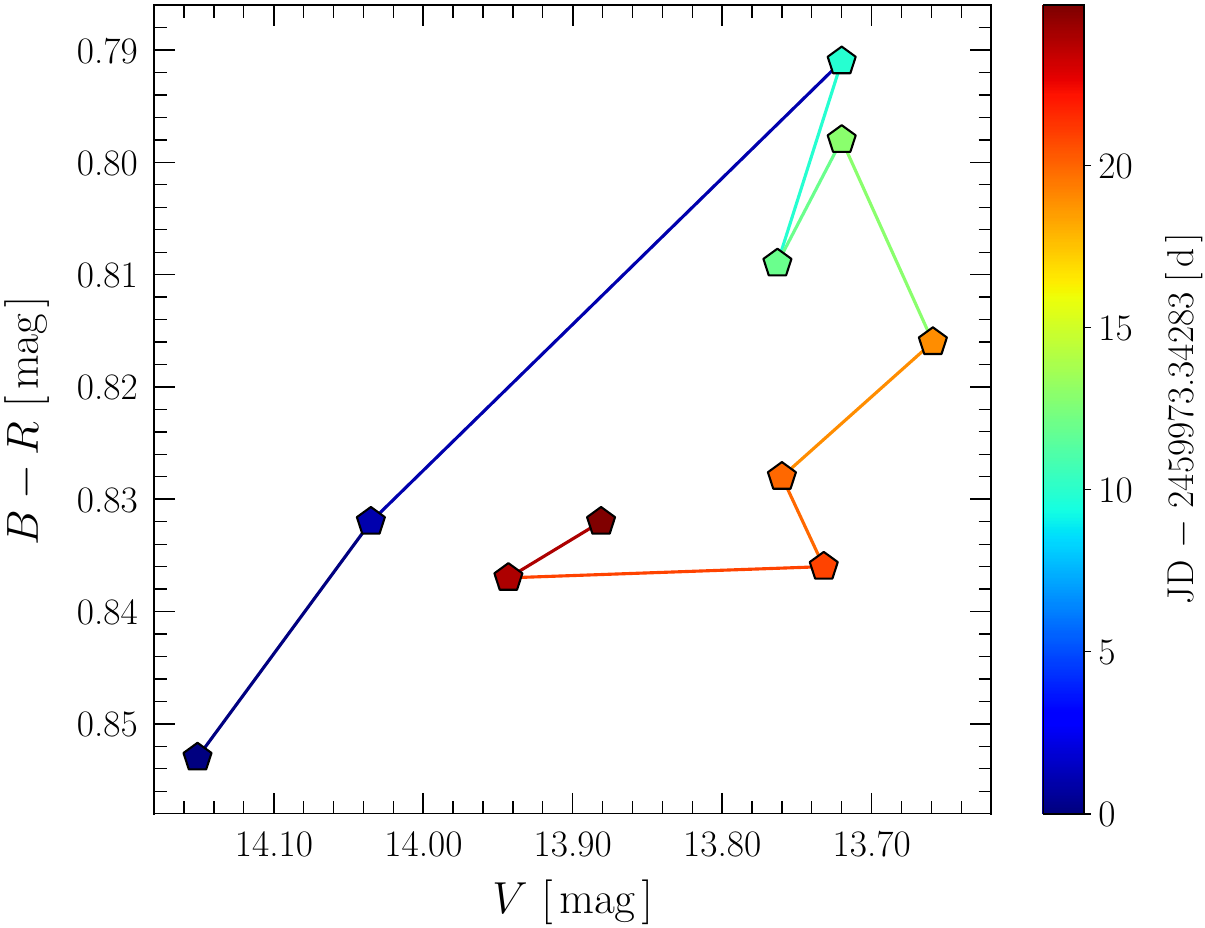}
    \caption{Colour-magnitude diagram using only data points around JD 2459980. A clockwise spectral hysteresis loop can be traced. The different colours denote the different observing times as indicated on the right.}
    \label{fig:flare:cmd}
\end{figure}

We have shown that the two-blob model gives a quite reasonable description of the considered portion of the LC except for the region around JD 2459980.
The observed flux excess with respect to the two-blob model could be related to a synchrotron flare. In support of this possibility, an enhanced $\gamma$-ray emission is observed at the same time (see Fig.\,\ref{fig:lc:stv})~-- in the framework of the leptonic jet emission model, optical and $\gamma$-ray flares are produced by one and the same ensemble of electrons via synchrotron and inverse Compton radiation.
The evolution of the electron energy distribution owing to the acceleration and radiative cooling processes could lead to a spectral hysteresis loop as well. Its direction of rotation depends on the interplay between the acceleration, cooling, and light-crossing time-scales \citep{1999ASPC..159..325K}. To check for the loop existence, we built the colour-magnitude diagram using only data points around the putative synchrotron flare. The result is shown in Fig.\,\ref{fig:flare:cmd}, where a clockwise spectral hysteresis loop can be traced: the BWB and the redder-when-fainter trends form a crude ellipse.
In addition, we observed that the $R$-band LC around JD 2459980 tends to lag the $BV$-band ones (see Fig.\,\ref{fig:dcf:chu2} for $B$-band), which is expected if one has an evolution of the electron energy distribution owing to the acceleration and radiative cooling processes. 

Therefore, we consider a scenario that includes an interaction of the helically moving blobs with a jet substructure to explain the optical and $\gamma$-ray LCs around JD 2459980. This interaction results in an acceleration of electrons, which are then cooled via synchrotron optical and inverse-Compton $\gamma$-ray radiation, thus, producing simultaneous optical and $\gamma$-ray flares.

\begin{table}
    \centering
    \caption{Parameters of the two-blob model.}
    \label{tab:helix:fit}
    \begin{tabular}{cr@{$~\pm~$}lc}
    \hline
    Parameter & \multicolumn{2}{c}{Value} \\
    \hline
    $\zeta$         &   2\fdg2  & 0\fdg4 \\
    \noalign{\smallskip}
    $\psi_1$        &   2\fdg4  & 0\fdg8 \\
    $\varphi_{0,1}$ &  32\fdg6  & 3\fdg8 \\
    \noalign{\smallskip}
    $\psi_2$        &   0\fdg15 & 0\fdg03 \\
    $\varphi_{0,2}$ & 134\fdg9  & 18\fdg6 \\
    \hline
    \end{tabular}
\end{table}

\begin{figure}
    \includegraphics[width=1.0\columnwidth,clip=true]{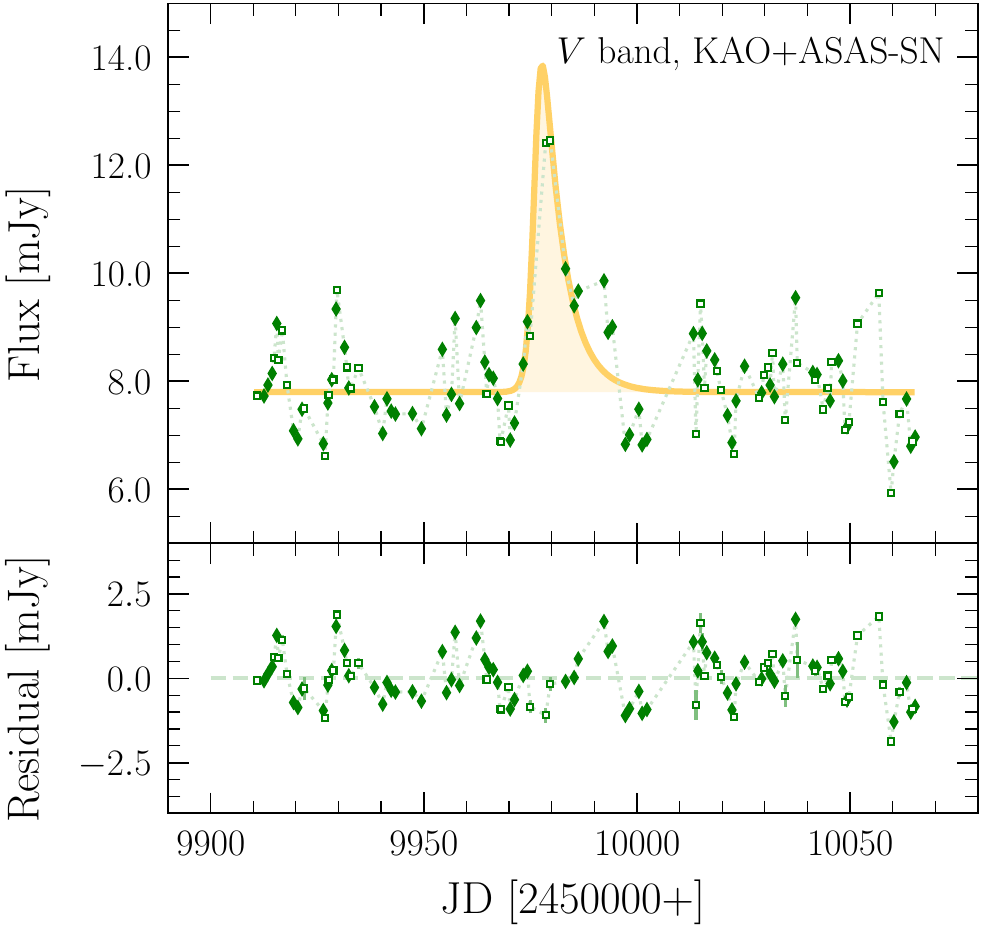}
    \caption{Deboosted combined $V$-band LC (top panel; the symbols are the same as in Fig.\,\ref{fig:lc:stv}). The synchrotron flare fitted to the LC is marked by a solid line; the fitting residuals are shown in the bottom panel.}
    \label{fig:flare:fit}
\end{figure}

Using the two-blob model LC, we corrected the observed LC for the Doppler factor variations \citep[i.e. we performed deboosting;][]{2002A&A...390..407V,2004A&A...421..103V,2023ApJS..265...51A}. The resulting LC is shown in Fig.\,\ref{fig:flare:fit}: the main feature is the synchrotron flare. It was fitted with a double exponential law \citep{2010ApJ...722..520A}:
\begin{equation}
    F(t) = F_{\rm base} +
    2F_0 \left[\exp\left(\frac{t_0-t}{\mathcal{T}_{\rm r}}\right)+\exp\left(\frac{t-t_0}{\mathcal{T}_{\rm d}}\right)\right]^{-1},
\end{equation}
where $F_{\rm base}$ is the constant baseline level, $F_0$ is the flare amplitude (with respect to the baseline level), $t_0$ is the position in time of the symmetric flare peak, and $\{\mathcal{T}_{\rm r},\mathcal{T}_{\rm d}\}$ are the flare $e$-folding rise and decay time-scales. We assumed ${\cal T_{\rm d}} = \varpi\,{\cal T_{\rm r}}$, where $\varpi \ge 1$. If the flare is asymmetric, then (i) the time of the flare maximum, $t_{\rm max}$, differs from $t_0$ and (ii) the asymmetry parameter, $\xi$, could be introduced \citep[see][for relevant formulae]{2010ApJ...722..520A}; the total duration of the flare is $\Delta t \simeq 2({\cal T_{\rm r}} + {\cal T_{\rm d}})$.
We have deboosted the LC and, so, the baseline level was set to the minimal value of the two-blob model fit, $F_{\rm base}=7.8$\,mJy, and it was held fixed during the fitting. 
The flare fit is shown in Fig.\,\ref{fig:flare:fit}, and the best-fitting flare parameters are listed in Table\,\ref{tab:flare:fit}.

We related the rise time-scale to the radius, $\mathcal R$, of the emitting region under the assumption that the rest-frame electron injection time is shorter than the light crossing time, $t'_{\rm inj} \lesssim t'_{\rm cros} \lesssim {\mathcal T_{\rm r}}\, \delta / (1+z)$:
\begin{equation} 
    {\mathcal R} \lesssim {\mathcal R}_{\rm max} = c\,{\mathcal T_{\rm r}} \left(\frac{\delta}{1+z}\right) \quad \rm [cm];
\end{equation}
here and below the time-scales are in seconds.

The decay time-scale is an upper limit of the rest-frame synchrotron cooling time of the emitting electrons, $t'_{\rm cool} \lesssim {\mathcal T_{\rm d}}\, \delta / (1+z)$. Taking into account the expression for $t'_{\rm cool}$ and the relation between the frequency of the emitted synchrotron radiation and the Lorentz factor of the emitting electrons, $\gamma$, we can set limits on the rest-frame magnetic field strength, ${\mathcal B}$, and $\gamma$ \citep[e.g.][]{1997A&A...327...61G,2013ApJ...764...57D,2015A&A...578A..68C,2021RAA....21..302F,2023ApJS..265...51A}:
\begin{equation} 
    {\mathcal B} \gtrsim {\mathcal B}_{\rm min} \simeq 1307\, {\mathcal T_{\rm d}}^{\,-2/3}\, (1+q)^{\,-2/3}\, \left(\frac{1+z}{\delta\,\nu_{15}}\right)^{1/3} \quad [\rm G],
\end{equation}
\begin{equation} 
    \gamma \lesssim \gamma_{\rm max} \simeq 453\, \nu^{2/3}_{15}\left[\frac{{\mathcal T_{\rm d}}\, (1+q)\, (1+z)}{\delta}\right]^{1/3},
\end{equation}
where $\nu_{15}$ is the observed photon frequency (in units of $10^{15}$\,Hz) and $q$ is the Compton dominance parameter. The latter is the ratio of the co-moving energy densities of the radiation and magnetic fields, $q=U'_{\rm rad}/U'_{\mathcal B}$, and is set to $q = 0.3$ \citep{2010ApJ...716...30A,2017ApJ...841..113M}.
The so-derived limits on the parameters, namely radius, magnetic field strength, and electron Lorentz factor of the region responsible for the synchrotron flare are:
\begin{align}
    \nonumber
    {\mathcal R}_{\rm max} &\simeq (2.1\,\delta)\times10^{15} \simeq 3.3\times10^{16}\,\rm cm, \\
    \nonumber
    {\mathcal B}_{\rm min} &\simeq \frac{0.7}{\delta^{1/3}} \simeq 0.3\,\rm G, \\
    \nonumber
    \gamma_{\rm max} &\simeq \frac{15776}{\delta^{1/3}} \simeq 6314;
\end{align}
we used $\delta = 15.6$ \citep[a weighted mean over the values reported in][]{2017ApJ...846...98J}.
The obtained upper limit on the radius of the emitting region is comparable to the typical values used in the SED modelling of \sfive\ \citep[10$^{16}$--10$^{17}$\,cm;][]{2014ApJ...783...83L,2018A&A...619A..45M,2024PASA...41..103B,2025ApJ...980...19O}.

\begin{table}
    \centering
    \caption{Parameters of the synchrotron flare.}
    \label{tab:flare:fit}
    \begin{tabular}{cr@{$~\pm~$}lcc}
    \hline
    Parameter & \multicolumn{2}{c}{Value} & Unit \\
    \hline
    $t_0$              & 9976.44 & 0.19 & JD -- 2450000 \\
    $F_0$              &    4.79 & 0.17 & mJy \\
    \noalign{\smallskip}
    ${\cal T}_{\rm r}$ &  1.01 & 0.09 & d \\
    $\varpi$           &  4.86 & 0.50 &   \\
    ${\cal T}_{\rm d}^{\,(a)}$   & 4.91 & 0.67 & d \\
    $\Delta{\cal T}$   & 11.84 & 1.35 & d \\
    \noalign{\smallskip}
    $\xi$              &    0.66 & 0.05 &    \\
    $t_{\rm max}$      & 9977.76 & 0.24 & JD -- 2450000 \\
    \hline
    \end{tabular}
    \flushleft{{\em Note.} We obtained $F(t_{\rm max}) \simeq 13.83$\,mJy by means of an interpolation. \\ $^{(a)}$ Computed as ${\cal T_{\rm d}} = \varpi\,{\cal T_{\rm r}}$.}
\end{table}

\subsection{Intranight variability}
\label{sec:dc:d}

Our INM resulted in a very low DC (about 10--20 per cent; see Sect.\,\ref{sec:inv}), which is unusual for \sfive; the typical values are larger than 50 per cent. \citet{2024MNRAS.527.5220T} even considered the \sfive\ DCs, derived by them (58--75 per cent depending on the band), lower than those obtained using a combined sample of intranight LCs for over the last three decades.

Nevertheless, low DC values ($\lesssim$30 per cent) were already reported in the literature: 
\citet{2015ApJS..218...18D} detected INV for 11 out of 72 nights (DC of $\sim$15 per cent);
\citet{2018AJ....155...31H} found a DC of 19.57 per cent (or 31.34 per cent if Var+PV cases are considered);
\citet{2018AJ....156...36K} estimated the DC to be 31 per cent using the nights with monitoring duration of two hours at least.
The DC derived in this work is comparable to or lower than the above-mentioned values.

The duration of our INM sessions is generally longer than $\sim$4\,h (except in two occasions) and so, the short time span should not be the cause of our low DC. 
If we do not consider the two cases of short INM duration, then the DC gets 11.1 per cent (or 22.1 per cent if we consider Var+PV cases), which is still a low value for \sfive.

The accuracy of the DC estimation could be improved by increasing the sampling of the blazar LC with INM runs. We supplemented our data with the results of \citet{2024MNRAS.527.5220T} and \citet{2024ApJ...971...74E}.
These authors monitored \sfive\ for more than two hours during three nights in 2022 December, which falls within our observing campaign. Using a total of 14 nights of INM, we got a DC of 8.1 per cent for the period of our observing campaign (or 16.1 per cent if we consider Var+PV cases), that is, the DC gets even lower.
If we further include the 2022 October data of the above authors~-- three more nights, which are still close in time to the beginning of our campaign~-- then the DC decreases down to 7.2 per cent (or to 14.4 per cent if we consider Var+PV cases).
The low DC values obtained by us may signal the need to consider the time dependence of the intranight DC in a more sophisticated way.

As noted in Sect.~\ref{sec:inv}, no flares are observed.
In the framework of the turbulent jet model, flares result from interactions of shocks with small-scale structures (e.g. density inhomogeneities, turbulent cells) in the jet flow \citep[e.g.][]{2013A&A...558A..92B,2023ApJS..265...51A}. Therefore, the lack of flares on intranight time-scales indicates shock propagation through a relatively homogeneous jet medium \citep[e.g.][]{2017Galax...5...77W}.

The formation of inhomogeneities could be related to the Kelvin-Helmholtz instability. It occurs at the boundary of two jet components with different velocities and densities and could give rise to considerable changes in the jet flow morphology. However, if the magnetic field is strong enough, ${\mathcal B} > {\mathcal B}_{\rm crit}$, then the Kelvin-Helmholtz instability cannot develop and, hence, the flaring intranight activity ceases. The critical magnetic field value is \citep{1995Ap&SS.234...49R}:
\begin{equation}
    {\mathcal B}_{\rm crit} = \frac{1}{\Gamma} \sqrt{~4 \pi n_{\rm e} m_{\rm e} c^2 \left(\Gamma^{\,2} - 1\right) ~} \quad \rm [G],
\end{equation}
where $\Gamma$ is the bulk Lorentz factor of the flow, $n_{\rm e}$ is the local electron density, and $m_{\rm e}$ is the rest mass of the electron; we assumed $\Gamma = 14.0$, while $n_{\rm e} = 23.85\,\rm cm^{-3}$ was taken from \citet{2025ApJ...980...19O}. Thus, we derived ${\mathcal B}_{\rm crit} \simeq 0.02$\,G. The recent magnetic field strength estimates during the \sfive\ low state in 2022 March$-$April by means of SED modelling resulted in ${\mathcal B} \simeq 0.6\,\rm G$ \citep{{2024PASA...41..103B}}. This value is larger than the above estimate of ${\mathcal B}_{\rm crit}$, that is, the formation of inhomogeneities by the Kelvin-Helmholtz instability is suppressed.

\section{Summary}
\label{sec:sum}

The main results of this study, according to the time-scales of the flux variations, could be summarized as follows. 

On long-term time-scales:
\begin{enumerate}
    \item The variability amplitude for the combined $V$-band LC for the period from 2012 January 24 to 2023 May 28 is estimated to $388.50 \pm 0.08$ per cent.
    \item The activity of \sfive\ shows gradual decrease. In the optical, the variability amplitude decreases down to $97.59 \pm 0.02$ per cent (calculated only for the period of our observing campaign). In the $\gamma$-rays, the fractional variability decreases from 0.64 (before JD 2459000) to 0.49 (after JD 2459000).
    \item We speculate that there are two episodes of new jet component ejection, not associated with $\gamma$-ray flares in this low-activity period.
\end{enumerate}

On short-term time-scales:
\begin{enumerate}
    \item  The percentage variability amplitude is $97.59 \pm 0.02$ per cent in the $V$ band and is found to decrease from $B$ to $R$ bands. A moderate FWB spectral behaviour is present. The strength of the anti-correlation between the spectral index and the flux gets weaker as the wavelength increases. The weighted mean spectral index is estimated to $1.139 \pm 0.004$.
    \item The $BVR$-band SF slopes are consistent with that of the red-noise. For all bands the time asymmetry of the LCs is negative, that is, we have shallower flux rises than decays.
    \item The flux variations in the $BVR$ bands are strongly correlated with no time lags among them.
    \item Using various techniques, we detected a transient QPO with a period of $43.5 \pm 3.6$\,d. This QPO was interpreted in terms of helical motion of two blobs within the jet; we derived the corresponding model parameters.
    \item The aperiodic signal detected in the LC was interpreted in terms of electron acceleration and cooling; the parameters of the resulted synchrotron flare are estimated. In addition, we derived limits on the radius, magnetic field strength, and electron Lorentz factor of the region related to the synchrotron flare.
\end{enumerate}

On intranight time-scales:
\begin{enumerate}
    \item We estimated INV DC of 10.3 per cent if we considered the PV case non-variable and 20.5 per cent if the PV case is variable; this DC is among the lowest one ever derived for \sfive.
    \item We observed smooth variability with no signs of flares. This could result from a temporarily homogeneous jet flow without formation of turbulent cells due to strong enough magnetic field preventing Kelvin-Helmholtz instability development. 
\end{enumerate}

In fine, \sfive\ manifested low activity on all time-scales studied by us.

\section*{Acknowledgements}

We thank the reviewer Prof. Junhui Fan for the valuable comments and suggestions that improved the paper. This work was supported by the Bulgarian Academy of Sciences under grant number IC-EG/09/2022-2024 and the Egyptian Academy of Scientific Research and Technology (ASRT) under grant number 10125. The project title is `Study of blazar jets through optical microvariability based on coordinated astronomical observations in Bulgaria and Egypt'.
The research that led to these results was carried out partly with the help of infrastructure renovated under the National Roadmap for Research Infrastructure (2020-2027), financially coordinated by the Ministry of Education and Science of Republic of Bulgaria.
This research has made use of the NASA/IPAC Extragalactic Database (NED), which is funded by the National Aeronautics and Space Administration and operated by the California Institute of Technology.
This research made use of {\sc ccdproc}, an {\sc astropy} package for image reduction \citep{matt_craig_2017_1069648}.
This study makes use of VLBA data from the VLBA-BU Blazar Monitoring Program (BEAM-ME and VLBA-BU-BLAZAR; http://www.bu.edu/blazars/BEAM-ME.html), funded by NASA through the {\em Fermi} Guest Investigator Program. The VLBA is an instrument of the National Radio Astronomy Observatory. The National Radio Astronomy Observatory is a facility of the National Science Foundation operated by Associated Universities, Inc.

\section*{Data Availability}

The data underlying this study will be available upon request to Dr. Ali Takey (ali.takey@nriag.sci.eg) or Prof. Boyko Mihov (bmihov@astro.bas.bg).



\bibliographystyle{mnras}
\bibliography{mnras0716} 




\appendix

\section{Intranight light curves}
\label{app:lc}

We present in Fig.\,\ref{fig:lc:inv} the intranight LCs obtained in the course of our observing campaign.

\begin{figure*}
\minipage{0.34\textwidth}
  \includegraphics[width=\columnwidth]{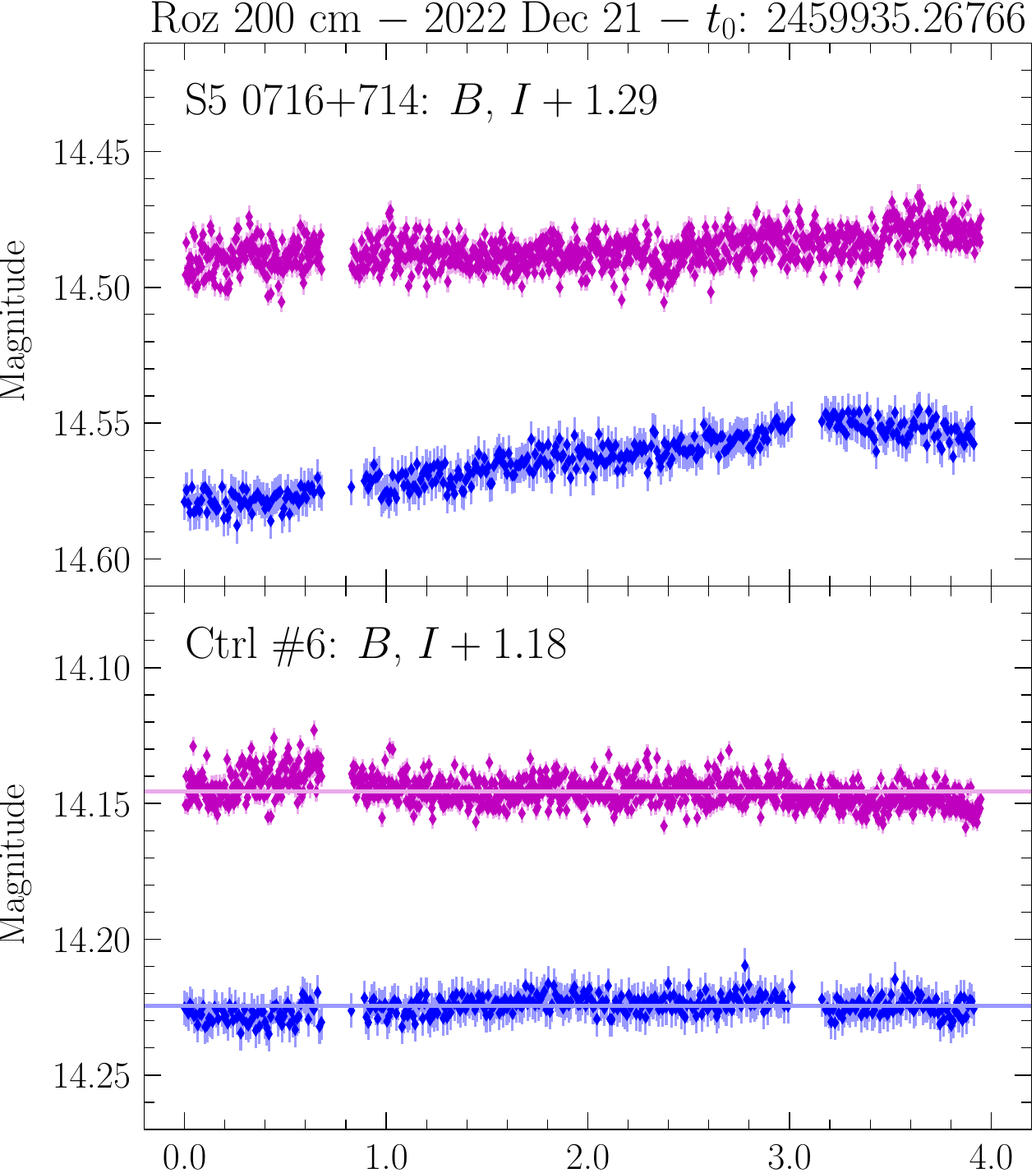}
\endminipage\hfill
\minipage{0.32\textwidth}
  \includegraphics[width=\columnwidth]{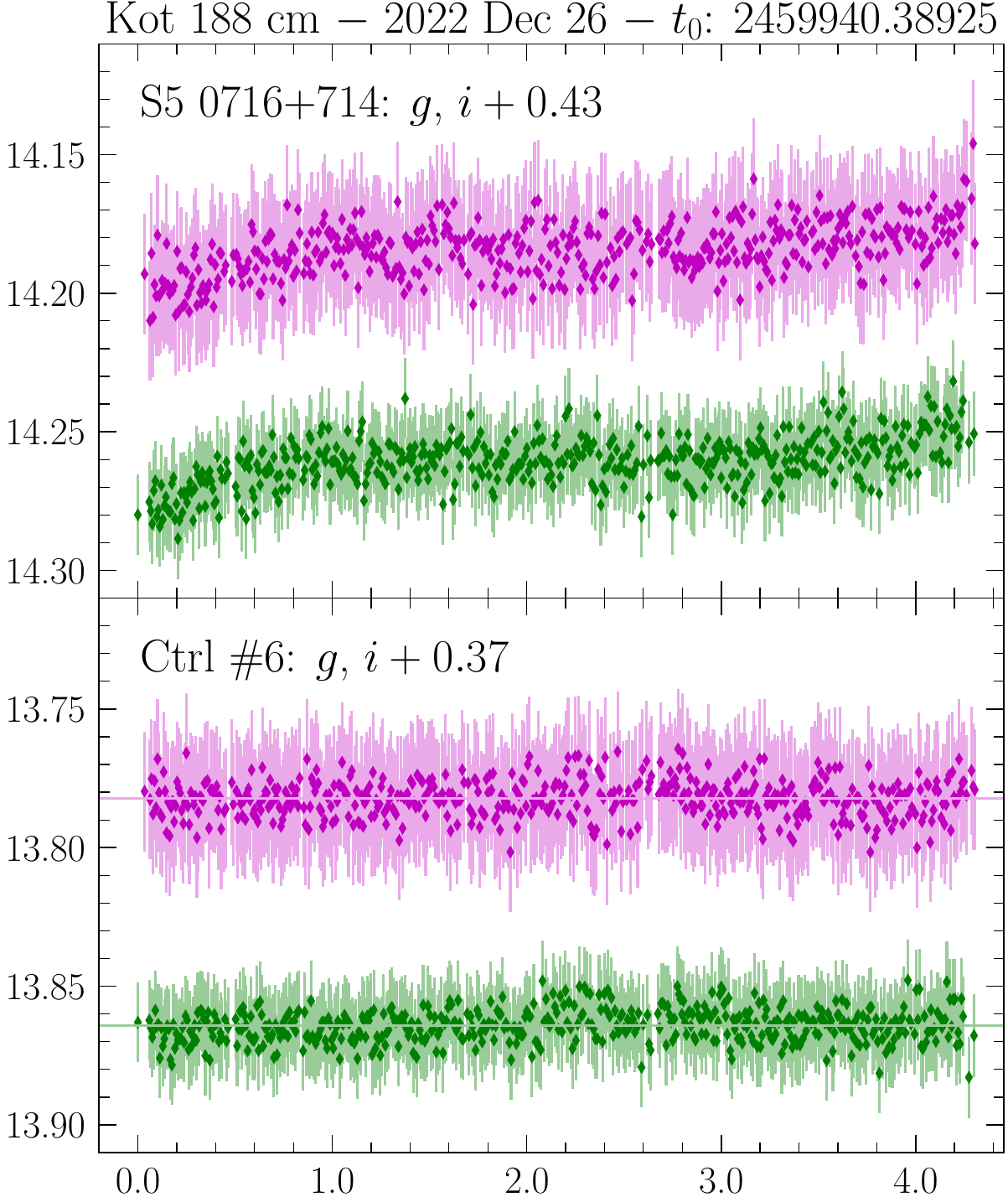}
\endminipage\hfill
\minipage{0.32\textwidth}
  \includegraphics[width=\columnwidth]{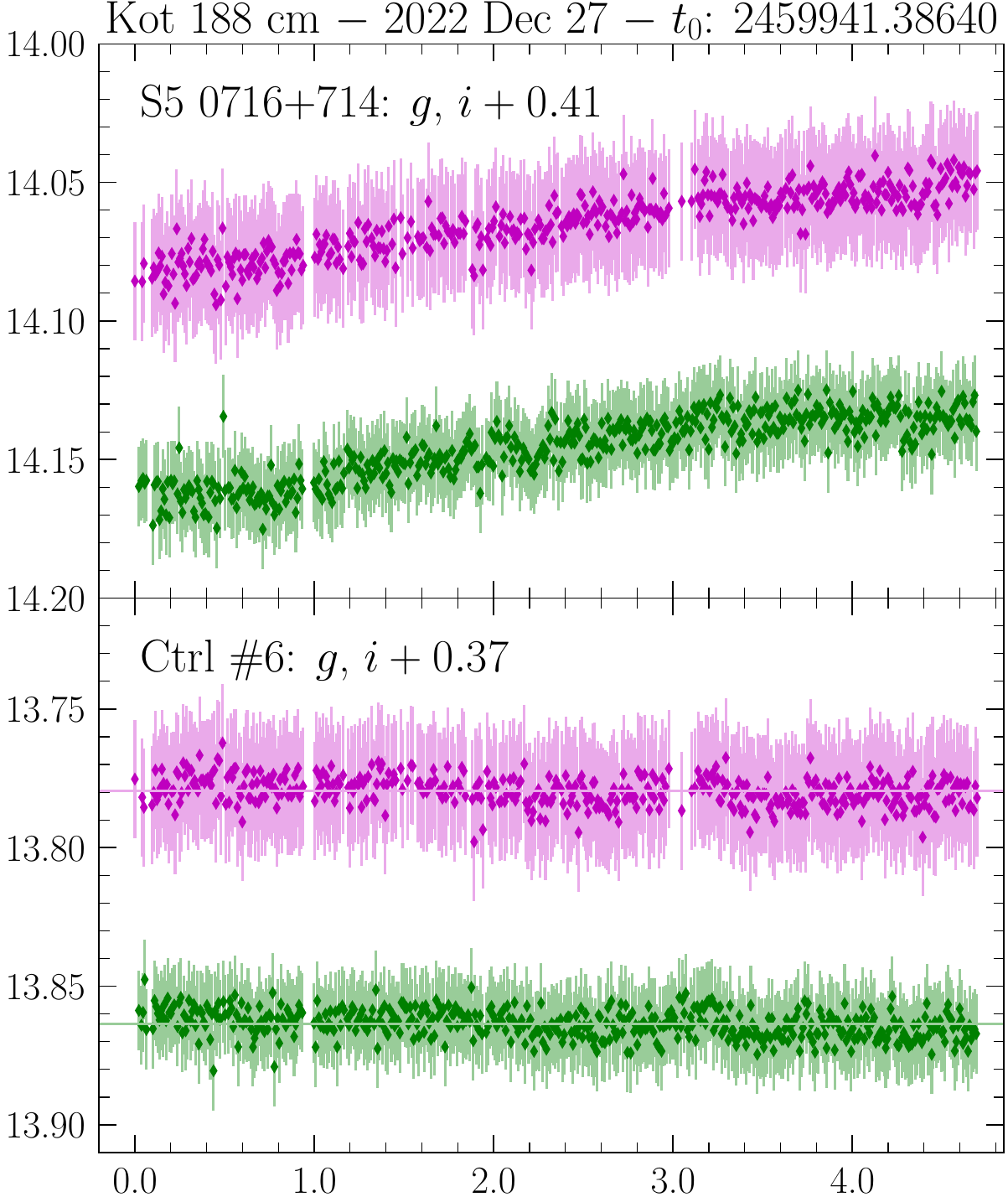}
\endminipage\hfill
\vspace{0.1cm}
\minipage{0.34\textwidth}
  \includegraphics[width=\columnwidth]{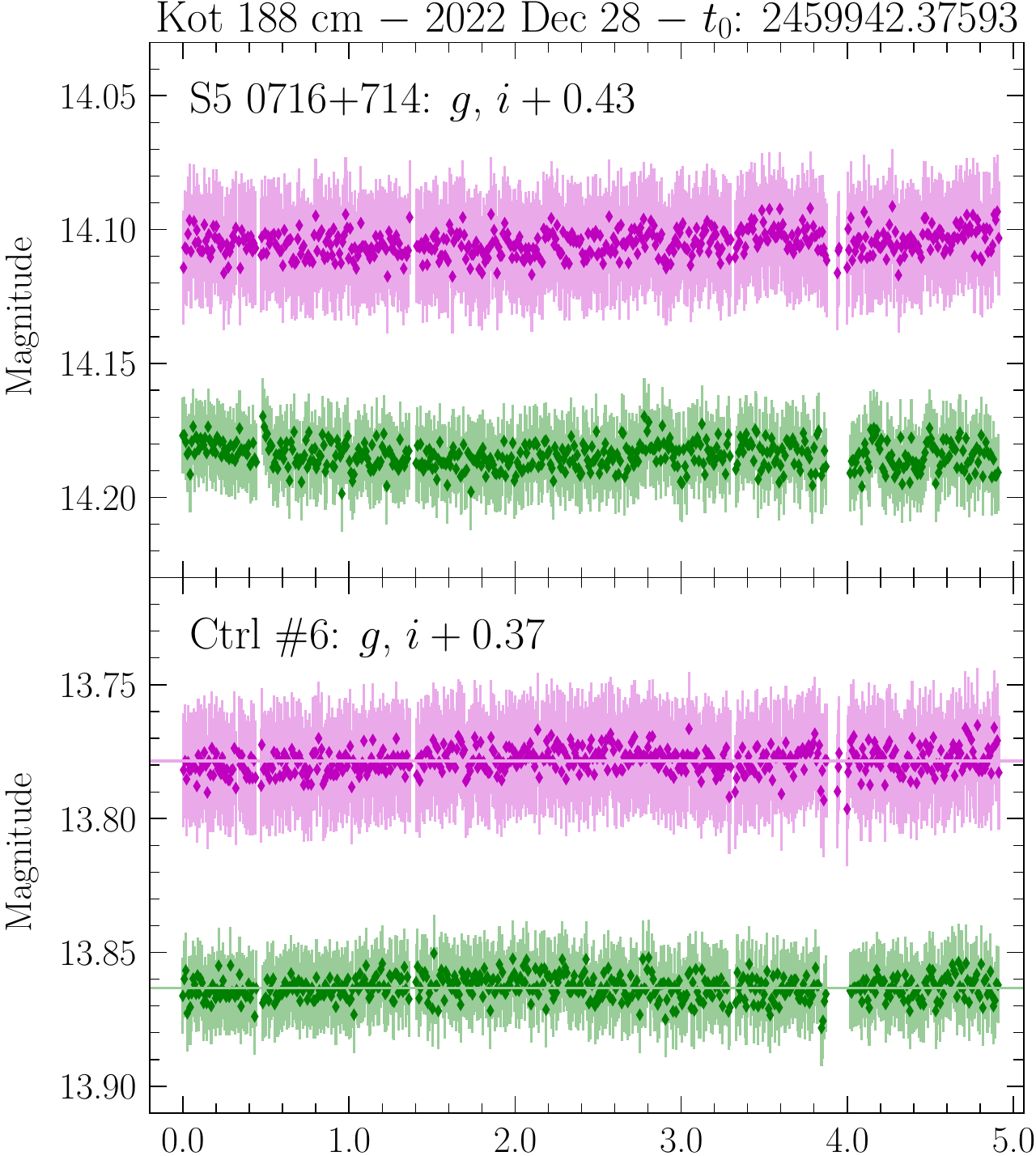}
\endminipage\hfill
\minipage{0.32\textwidth}
  \includegraphics[width=\columnwidth]{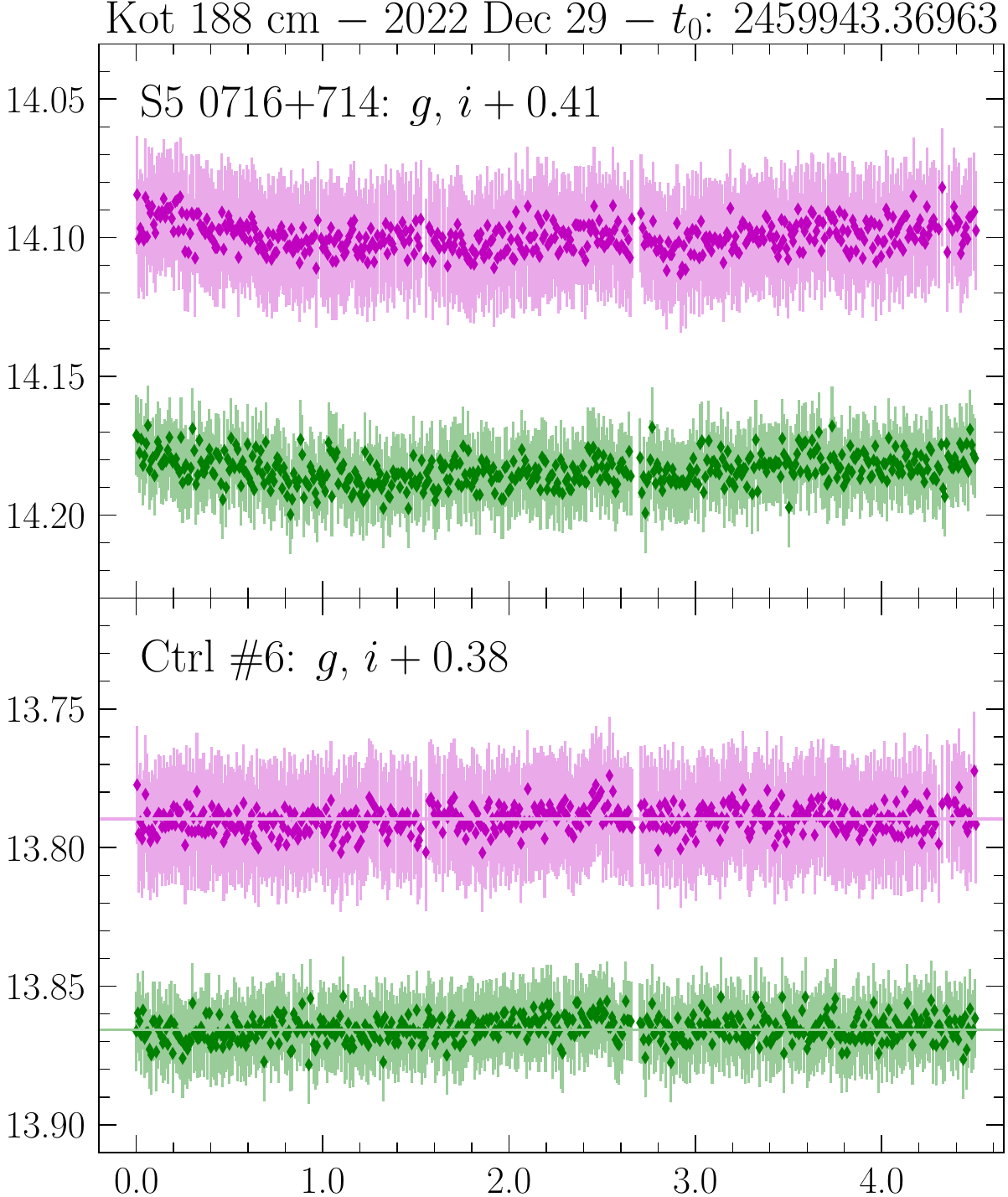}
\endminipage\hfill
\minipage{0.32\textwidth}
  \includegraphics[width=\columnwidth]{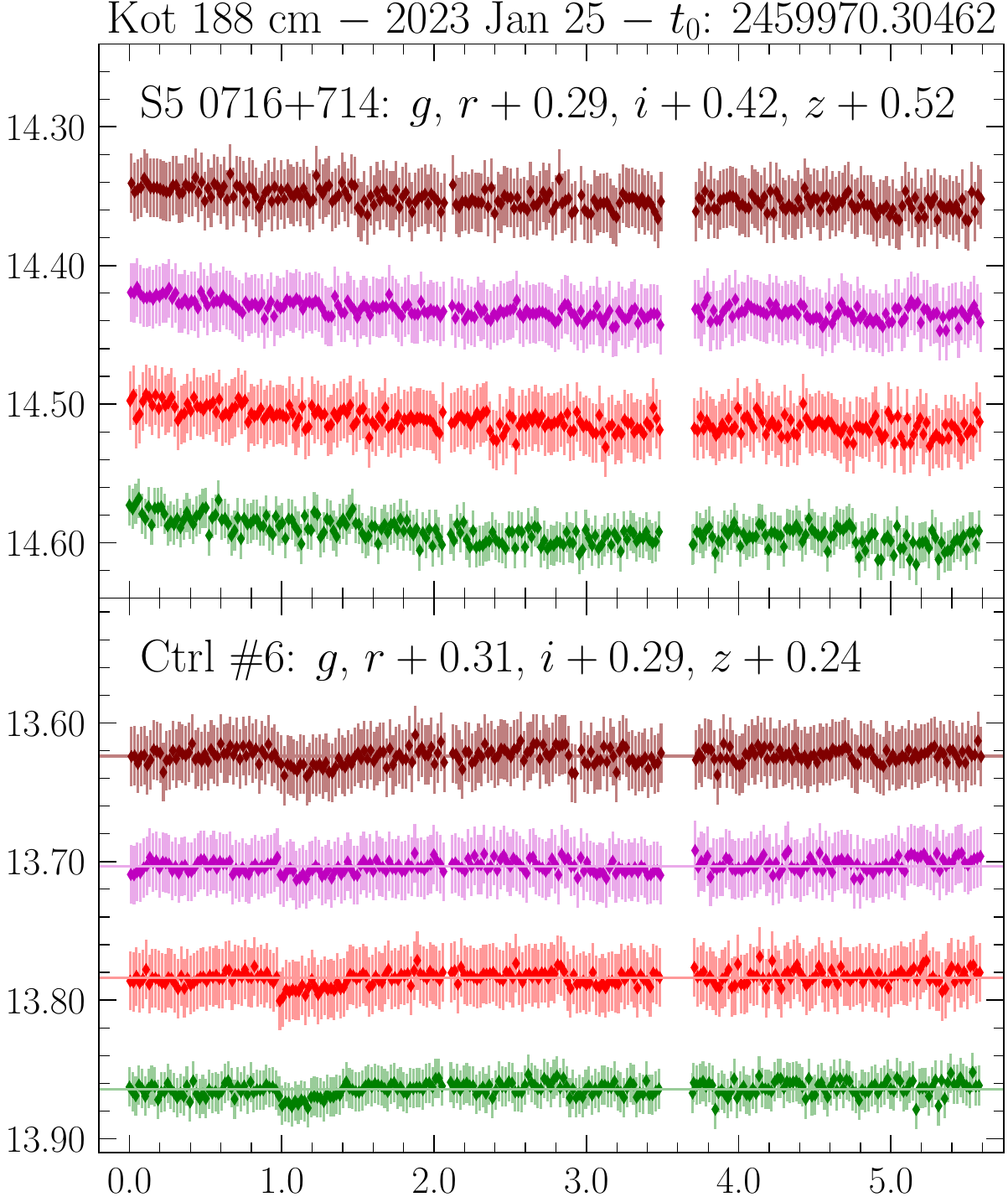}
\endminipage\hfill
\vspace{0.1cm}
\minipage{0.34\textwidth}
  \includegraphics[width=\columnwidth]{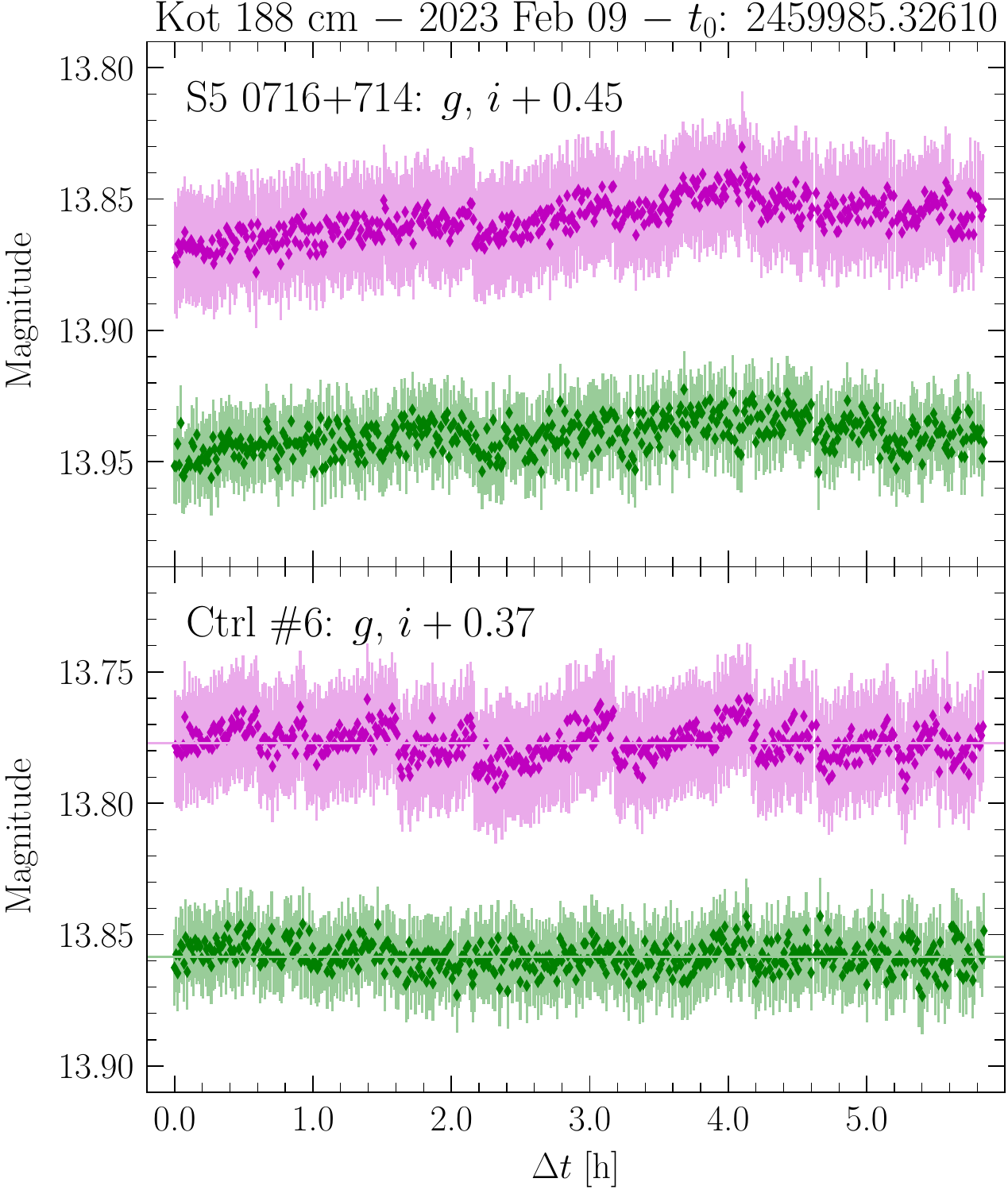}
\endminipage\hfill
\minipage{0.32\textwidth}
  \includegraphics[width=\columnwidth]{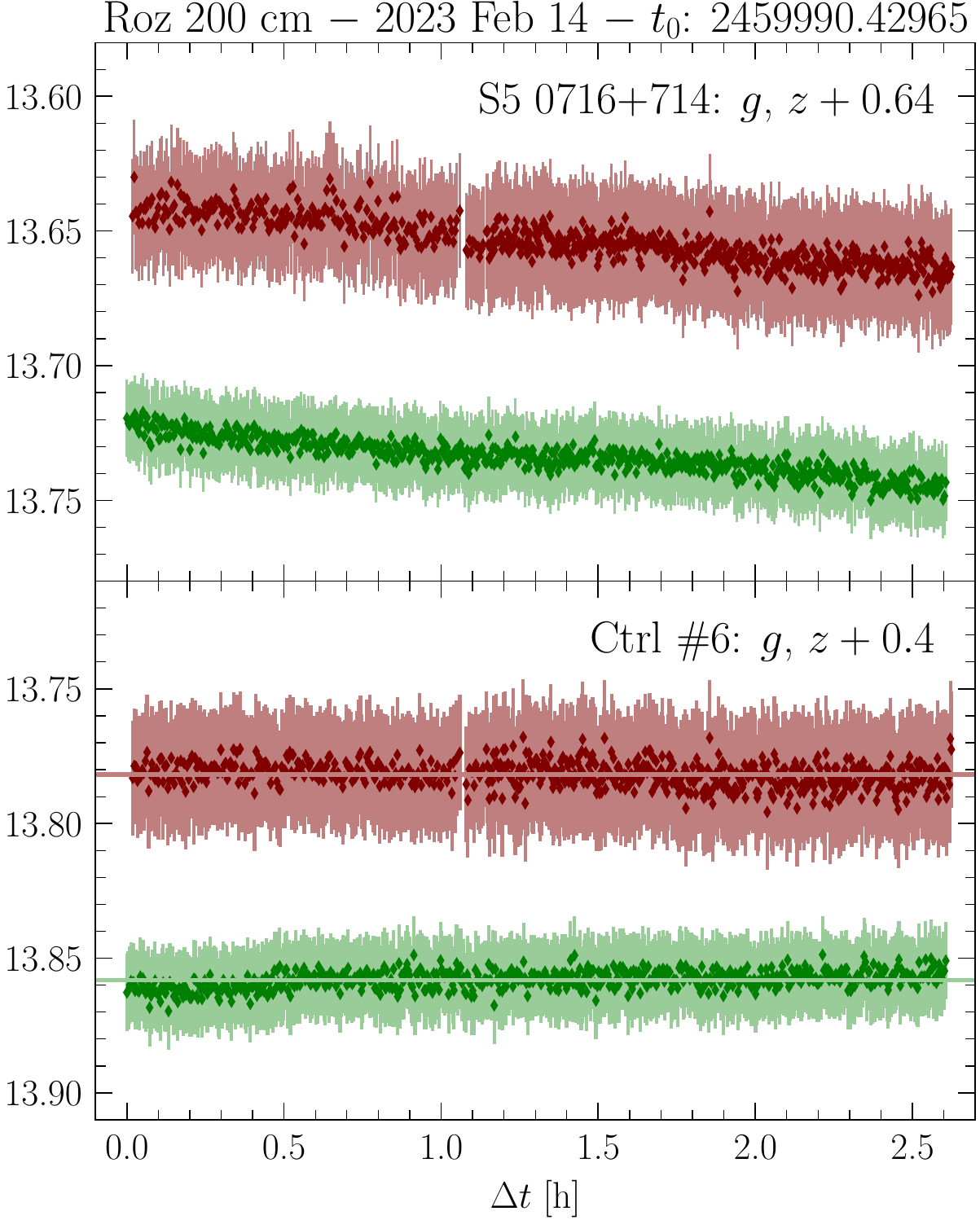}
\endminipage\hfill
\minipage{0.32\textwidth}
  \includegraphics[width=\columnwidth]{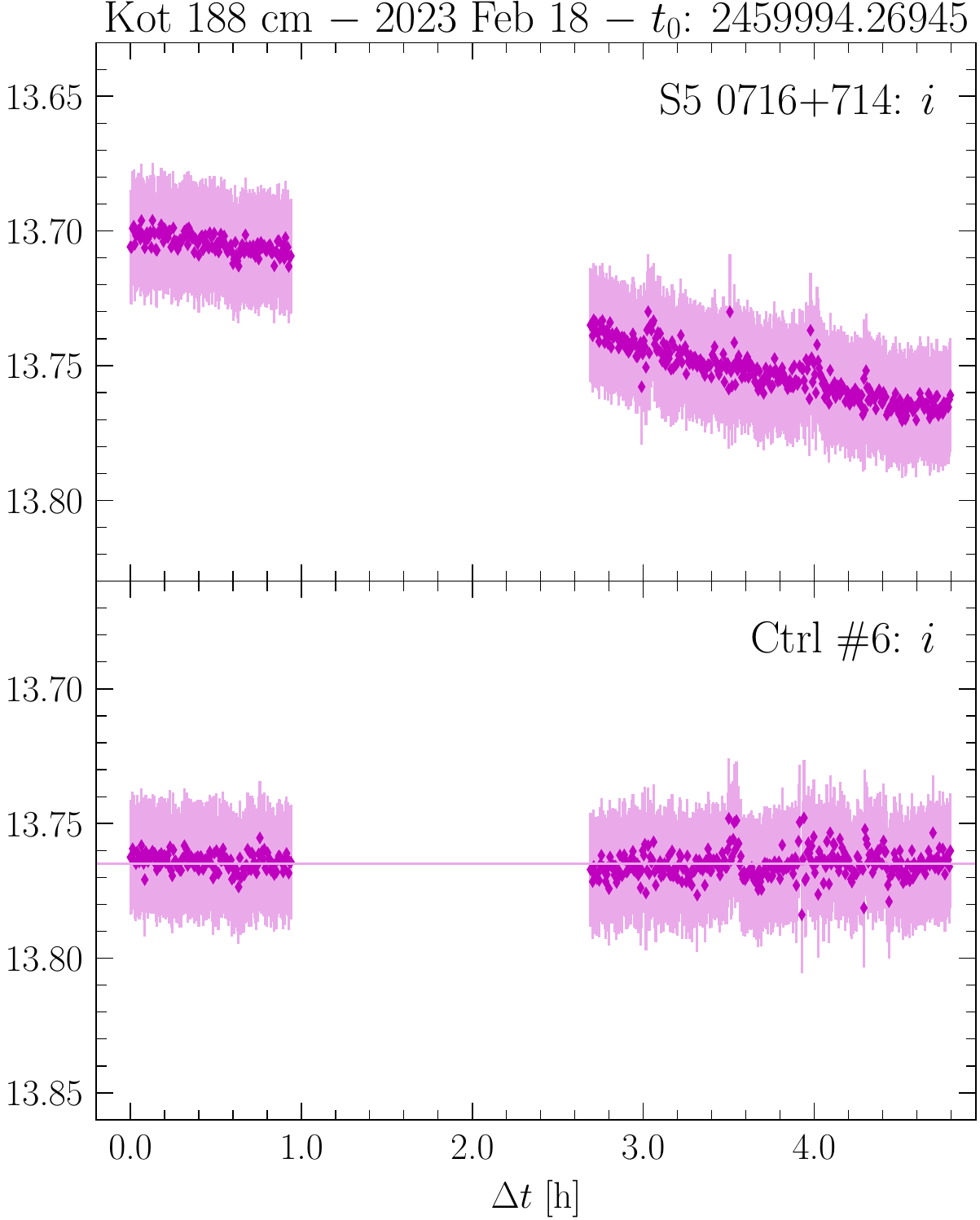}
\endminipage\hfill
\caption{Intranight LCs for \sfive. The telescope, evening date, starting epoch of the INM $t_0$, and the offsets applied for a better display of the LCs, are indicated in each plot. The LCs are coloured as follows: $B$~-- blue, $V$/$g$~-- green, $R$/$r$~-- red, $I$/$i$~-- magenta, $z$~-- dark red. They are ordered from bottom to top starting with the bluest one. The horizontal lines mark the weighted mean magnitudes of the control star.}
\label{fig:lc:inv}
\end{figure*}

\addtocounter{figure}{-1}
\begin{figure*}
\includegraphics[width=0.34\textwidth,clip=true]{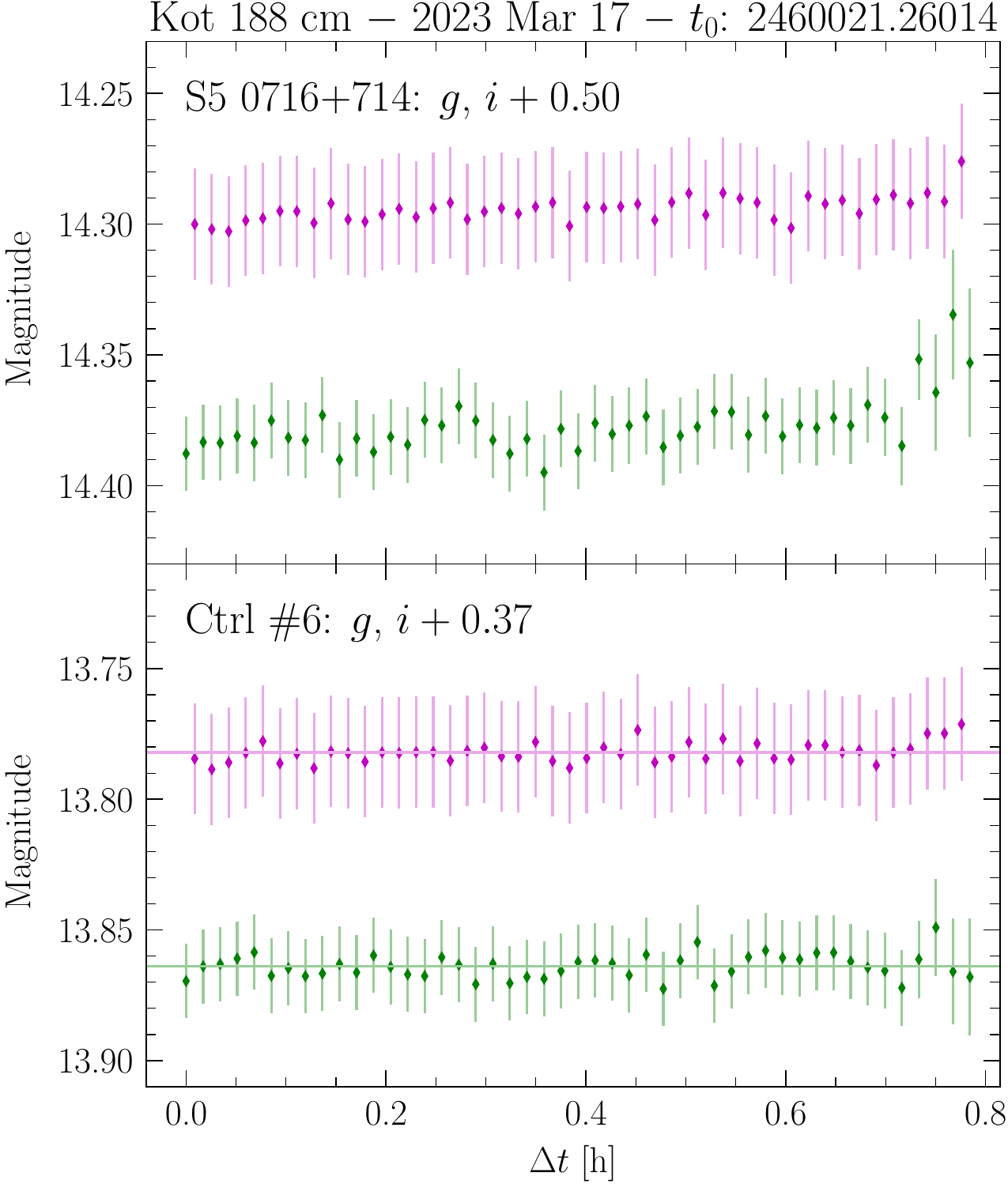}
\includegraphics[width=0.32\textwidth,clip=true]{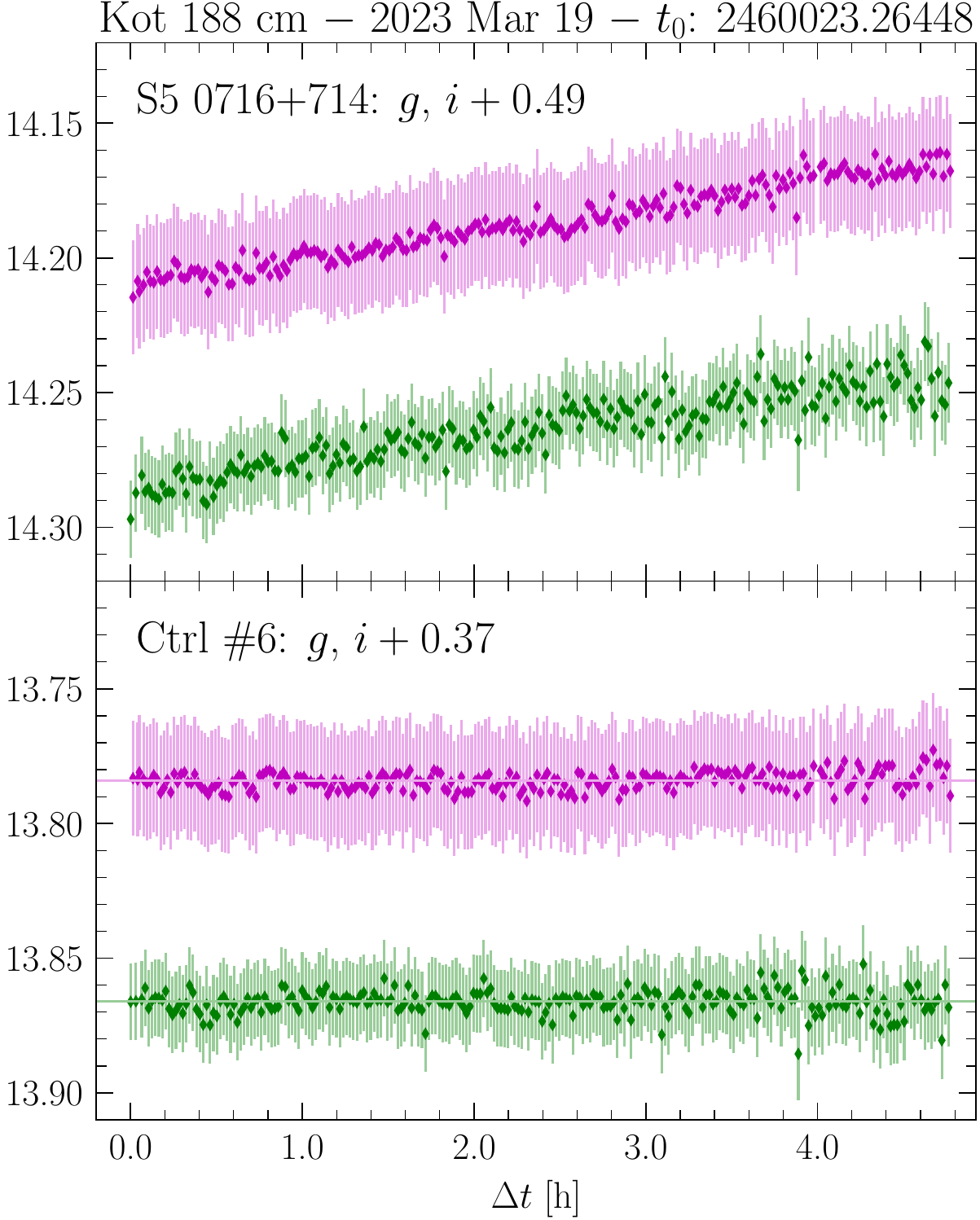}
\caption{Continued.}
\end{figure*}


\bsp	
\label{lastpage}
\end{document}